\documentclass[aps,prx,twocolumn,superscriptaddress]{revtex4-1}
\pdfoutput=1
\usepackage{amsmath}
\usepackage{amssymb}
\usepackage{graphicx}
\usepackage{bm}
\usepackage{hyperref}

\begin{document}

\title{Propulsion of a domain wall in an antiferromagnet by magnons}

\author{Se Kwon Kim}
\affiliation{Department of Physics and Astronomy, 
	The Johns Hopkins University,
	Baltimore, Maryland 21218, USA
}

\author{Yaroslav Tserkovnyak}
\affiliation{Department of Physics and Astronomy, 
	University of California, 
	Los Angeles, California 90095, USA
}

\author{Oleg Tchernyshyov}
\affiliation{Department of Physics and Astronomy, 
	The Johns Hopkins University,
	Baltimore, Maryland 21218, USA
}

\begin{abstract}
We analyze the dynamics of a domain wall in an easy-axis antiferromagnet driven by circularly polarized magnons. Magnons pass through a stationary domain wall without reflection and thus exert no force on it. However, they reverse their spin upon transmission, thereby transferring two quanta of angular momentum to the domain wall and causing it to precess. A precessing domain wall partially reflects magnons back to the source. The reflection of spin waves creates a previously identified reactive force. We point out a second mechanism of propulsion, which we term redshift: magnons passing through a precessing domain wall lower their frequency by twice the angular velocity of the domain wall; the concomitant reduction of magnons' linear momentum indicates momentum transfer to the domain wall. We solve the equations of motion for spin waves in the background of a uniformly precessing domain wall with the aid of supersymmetric quantum mechanics and compute the net force and torque applied by magnons to the domain wall. Redshift is the dominant mechanism of propulsion at low spin-wave intensities; reflection dominates at higher intensities. We derive a set of coupled algebraic equations to determine the linear velocity and angular frequency of the domain wall in a steady state. The theory agrees well with numerical micromagnetic simulations.
\end{abstract}

\maketitle

\section{Introduction}

Stability of domain walls and other topological defects makes them attractive candidates for use in technological applications exemplified by racetrack magnetic memory \cite{Science.320.190}. A major practical issue is finding a reliable means for moving domain walls. In a ferromagnet, an external magnetic field breaks the symmetry between domains with different orientations of magnetization and thereby applies a force to a domain wall. A spin-polarized electrical current has charge carriers adjusting their spins toward the local direction of magnetization and reacts by exerting a torque on a magnetic texture \cite{Berger1978, Slonczewski1996, Berger1996}. Developing basic models of the dynamics of topological defects in magnets is a major task for theorists. A classic example of such an effort is the 1974 paper of \textcite{Walker1973}, who successfully reduced a complex problem of magnetization dynamics near a domain wall to the evolution of its soft modes parametrized by two collective coordinates, position of the domain wall $X$ and azimuthal angle of magnetization $\Phi$. 

In this paper we present a theory of a domain wall propelled by spin waves in an antiferromagnet. A similar problem in a ferromagnet was analyzed by several groups \cite{Hinzke2011, Yan2011, Kovalev2012}. In that case, a magnon traversing a domain wall reverses its spin and deposits angular momentum $2\hbar$ on the domain wall. Addition of angular momentum to the domain wall translates directly into its shift toward the source of spin waves. The physics is different in an antiferromagnet in that translational motion of a wall is induced by transfer of linear momentum from magnons. \textcite{PhysRevLett.112.147204} found that circularly polarized spin waves propel a domain wall away from the source. Like in a ferromagnet, magnons deposit angular momentum $2\hbar$ on a domain wall. However, the addition of angular momentum does not translate directly into a displacement of the domain wall but rather causes it to precess. A precessing wall partially reflects spin waves back toward the source; \textcite{PhysRevLett.112.147204} inferred that the mechanism of propulsion was the reactive force of the reflected spin waves: upon reflection, a magnon with wavenumber $k$ alters its momentum from $+ \hbar k$ to $- \hbar k$ and thus transfers momentum $2\hbar k$ to the wall, pushing it away from the source. They argued for a steady state that a domain wall becomes a perfect reflector of spin waves, thus generating a maximal reactive force. 

\begin{figure}
\includegraphics[width=0.95\columnwidth]{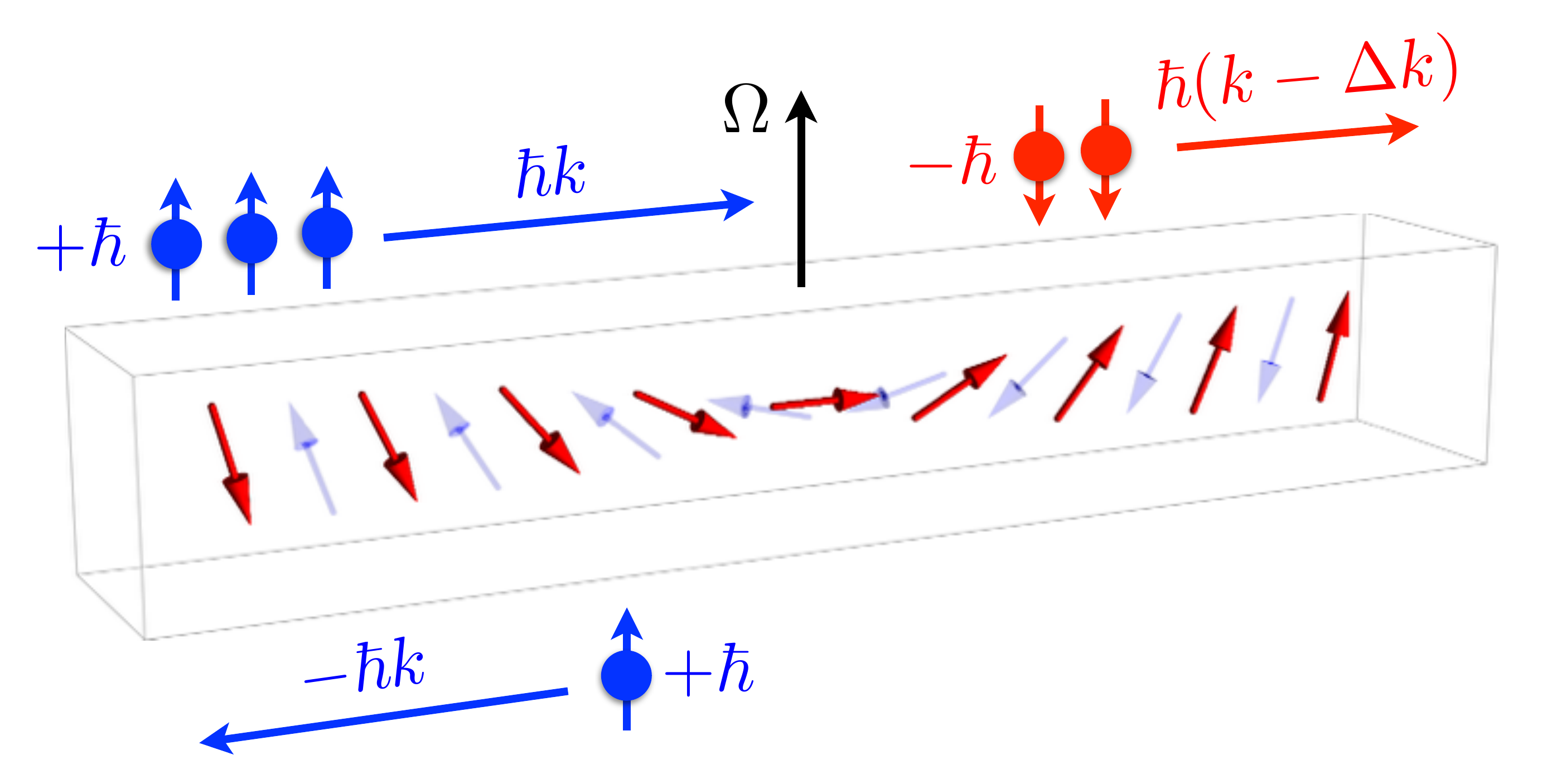}
\caption{Two mechanisms of domain-wall propulsion: reflection of spin waves by a precessing wall and redshift of the transmitted wave by twice the angular frequency of precession. In both cases, the change in magnon's linear momentum generates a reactive force on the domain wall. Spins on sublattices 1 and 2 are shown as solid red and faint blue arrows.}
\label{fig:mechanisms}
\end{figure}

Here we point out another mechanism of domain-wall propulsion by spin waves, which we term \emph{redshift} to distinguish it from the reactive force due to reflection. Magnons transmitted by a domain wall precessing at an angular velocity $\Omega$ experience a redshift in frequency by by $\Delta \omega = 2\Omega$. As a result, their momentum is reduced by $\hbar \Delta k = \hbar \Delta \omega/v_g$, where $v_g = d\omega/dk$ is the magnon group velocity. The missing momentum is transferred to the domain wall, which then accelerates. The two mechanisms are illustrated in Fig.~\ref{fig:mechanisms}. 

The relative importance of the reactive force components associated with magnon reflection and redshift depends on properties of a spin wave. Below we present a comprehensive analysis of this problem. It was facilitated by finding an exact solution for a spin wave in the background of a uniformly precessing domain wall, made possible by the use of supersymmetric quantum mechanics. The reactive force is dominated by magnon redshift at small amplitudes and by magnon reflection at large amplitudes. The crossover amplitude is determined by the wavevector of incoming magnons.

A suitable language for describing slow dynamics of a domain wall is the method of collective coordinates that has been developed for magnetic textures in ferromagnets \cite{Tretiakov2008} and in antiferromagnets \cite{Tretiakov2013}. The basic idea is to parametrize a magnetic texture using a set of collective coordinates $\{q_1, q_2, \ldots\}$. The kinetic energy of an antiferromagnet is expressed as $M_{ij} \dot{q}_i \dot{q}_j/2$, where $M_{ij}$ a mass tensor; the generalized (conservative) force conjugate to coordinate $q_i$ is obtained by differentiating potential energy, $F_i = - \partial U/\partial q_i$; the corresponding viscous force is $F^v_i = - D_{ij}\dot{q}_j$, where $D_{ij}$ is a dissipation tensor. The mass and dissipation tensors are proportional to each other, $D_{ij} = M_{ij}/T$; the relaxation time $T$ is inversely proportional to Gilbert's damping constant $\alpha$. (See Appendix~\ref{app:dynamics-general} for details.) The resulting equations of motion are in essence Newton's second laws for all the collective coordinates. To keep the problem tractable, one keeps only a small number of collective coordinates representing soft modes of the system. In our problem, we focus on the position of the the domain wall $X$ and its azimuthal angle $\Phi$ representing the soft modes associated with the symmetries of translation and spin rotation. Other variables---such as the width of the domain wall $\lambda$---represent hard modes and are assumed to adjust instantaneously to their equilibrium values \cite{Clarke2008}.

In this paper we follow a somewhat different approach and focus instead on two conserved quantities related to the symmetries of translation and spin rotation: linear momentum $P$ and angular momentum $J$ of the antiferromagnet. These physical variables have an intimate relation to the collective coordinates: they are the canonical momenta conjugate to the position of the domain wall $X$ and azimuthal angle $\Phi$. Whereas collective coordinates are convenient when forces acting on a magnetic texture can be encoded in a potential energy, in this paper we deal with reactive forces associated with reflection and transmission of spin waves. Forces of this kind are more easily computed in the language of conservation laws and conserved quantities. The two approaches can of course be combined to achieve greater clarity. 
 
The paper is organized as follows. The model and a summary of main results are outlined in Sec.~\ref{sec:summary}. In Sec.~\ref{sec:qft} we review the field theory of an antiferromagnet with easy-axis anisotropy and discuss the properties of stationary, moving, and precessing domain walls. In Sec.~\ref{sec:spin-waves} we obtain exact solutions for spin waves in the backgrounds of static and precessing domain walls with the aid of supersymmetric quantum mechanics \cite{JPhysA.18.2917, PhysRep.251.267}. In Sec.~\ref{sec:forces-torques} we derive the reactive force and torque exerted on a domain wall by spin waves and the viscous force and torque due to Gilbert damping. The resulting equations of motion for a domain wall are analyzed in Sec.~\ref{sec:analysis}. We conclude with a general discussion in Sec.~\ref{sec:discussion}.

\section{Summary of main results}
\label{sec:summary}

We consider an easy-axis antiferromagnet in one spatial dimension. Well below the ordering temperature, its staggered magnetization has a fixed length and can be encoded by the unit vector field $\mathbf n(x,t)$. In the continuum approximation, its dynamics are governed by kinetic and potential energy, whose densities are, respectively,
\begin{equation}
\mathcal K = \frac{\rho |\dot{\mathbf n}|^2}{2},
\quad
\mathcal U = \frac{A|\mathbf n'|^2 + K_0(\hat{\mathbf z} \times \mathbf n)^2}{2}.
\label{eq:K-U}
\end{equation}
Here $A$ is the exchange constant, $K_0>0$ is the strength of easy-axis anisotropy, and $\rho$ quantifies inertia of staggered magnetization. Invariance of the Lagrangian density $\mathcal L = \mathcal K - \mathcal U$ under spatial translations and under rotation of magnetic moments about the $z$-axis gives rise to conservation of linear momentum $P$ and angular momentum $J$.

The dynamics of the antiferromagnet has a ``relativistic'' form, with the role of the ``speed of light'' played by the maximal group velocity of spin waves $s = \sqrt{A / \rho}$. As a result, spin waves in a uniform ground state have a ``relativistic'' dispersion,
\begin{equation}
\omega^2 = \omega_0^2 + s^2 k^2,
\end{equation}
with the spin-wave gap $\omega_0 = \sqrt{K_0/\rho}$. On a deeper level, the equations of dynamics are invariant under ``Lorentz'' transformations, 
\begin{equation}
t \mapsto t' = \frac{t-vx/s^2}{\sqrt{1-v^2/s^2}},
\quad
x \mapsto x' = \frac{x-vt}{\sqrt{1-v^2/s^2}}.
\label{eq:x'-t'}
\end{equation}
It should be kept in mind that $t'$ and $x'$ are not physical time and coordinate in a moving frame (which would involve the speed of light $c$ instead of $s$) but are rather convenient formal variables that utilize the ``relativistic'' nature of magnetization dynamics in a uniaxial antiferromagnet. We will nonetheless refer to the pair $(t',x')$ as a moving reference frame. 

For convenience, we use natural units of length, time, and energy determined by the three coupling constant: 
\begin{equation}
\lambda_0 = \sqrt{A/K_0}, 
\quad
t_0 = \sqrt{\rho/K_0},
\quad 
\epsilon_0 = \sqrt{AK_0},
\end{equation}
which have transparent physical meaning. The width of a static domain wall is $\lambda_0$. The spin-wave frequency gap $\omega_0 = 1/t_0$. The energy of a static domain wall is $2 \epsilon_0$. 

\begin{figure}
\includegraphics[width=0.95\columnwidth]{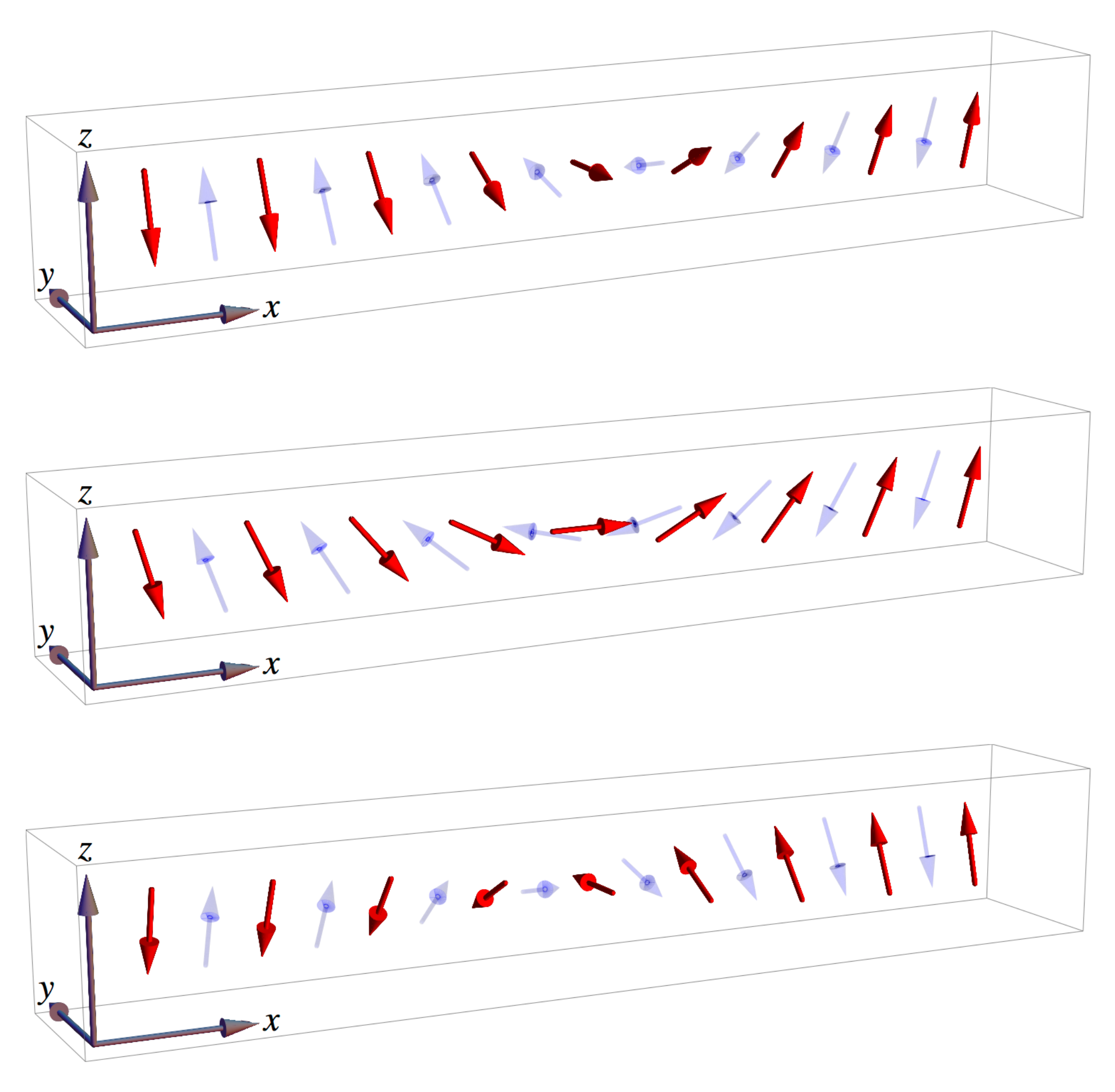}
\caption{Domain walls in an antiferromagnet (\ref{eq:domain-wall-static}) with different azimuthal angles $\Phi$. Top to bottom: $\Phi = -\pi/2$, 0, $+\pi/2$. Spins on sublattices 1 and 2 are shown as solid red and faint blue arrows.} 
\label{fig:domain-wall}
\end{figure}

The easy-axis antiferromagnet has two ground states with uniform staggered magnetization, $\mathbf n = \pm \hat{\mathbf z}$. \textcite{PhysRevLett.50.1153} obtained domain-wall (topological soliton) solutions between the two uniform ground states by minimizing the energy density $\mathcal H = \mathcal K + \mathcal U$ (\ref{eq:K-U}) with respect to staggered magnetization at fixed values of linear and angular momenta. A fixed domain wall that is rotating about the $z$-axis with the angular velocity $\Omega$ has the following profile: $\mathbf n_{\Omega} (t, x) = (\sin{\theta} \cos{\phi}, \, \sin{\theta} \sin{\phi}, \, \cos{\theta})$, where 
\begin{eqnarray}
&& \cos{\theta} = \tanh \left( x \sqrt{1 - \Omega^2} \right),
\nonumber \\
&& \sin{\theta} = \mathop{\mathrm{sech}} \left( x \sqrt{1 - \Omega^2} \right),
\nonumber \\
&& \phi = \Omega t.
\end{eqnarray}
We can obtain a moving domain-wall solution by ``Lorentz'' boosting a zero-velocity solution:
\begin{equation}
\mathbf n_{V, \Omega}(t, x) = \mathbf n_{\Omega} \left( \frac{t - V x}{\sqrt{1 - V^2}},  \frac{x - V t}{\sqrt{1 - V^2}} \right).
\end{equation}
The linear and angular momenta of the domain wall are related to its linear velocity $V$ and angular velocity $\Omega$ as follows:
\begin{subequations}
\begin{eqnarray}
P &=& \frac{M V}{\sqrt{ (1 - V^2) (1 - \Omega^2) }}, 
\\
J &=& \frac{I \Omega}{\sqrt{1 - \Omega^2}},
\end{eqnarray}
\end{subequations}
where $M = 2$ and $I = 2$ are the mass and moment of inertia of a static domain wall. For slow dynamics, $V \ll 1$ and $\Omega \ll 1$, we recover the non-relativistic relations, $P \approx M V$ and $J \approx I \Omega$.

The equations of motion for linear momentum $P$ and angular momentum $J$ in the presence of external force $F$ and torque $\tau$ read
\begin{subequations}
\begin{eqnarray}
\dot{P} &=& F + F^v ,
\\
\dot{J} &=& \tau + \tau^v,
\end{eqnarray}
\end{subequations}
where $F^v$ and $\tau^v$ are the viscous force and torque. It is convenient to work in the frame (\ref{eq:x'-t'}) moving at the instantaneous velocity of the domain wall $V$. In this frame, where the domain wall is fixed in space, 
\begin{subequations}
\begin{eqnarray}
F^v &=& - \frac{M V \sqrt{1 - \Omega^2}}{T \sqrt{1 - V^2}},
\\
\tau^v &=& - \frac{I \Omega}{T \sqrt{(1 - V^2)(1 - \Omega^2)}},
\end{eqnarray}
\end{subequations}
and $T$ is the relaxation time determined by the dissipation rate of energy. The equations for a steady state with constant linear velocity $V$ and angular velocity $\Omega$ are
\begin{subequations}
\begin{eqnarray}
F &=& \frac{M V \sqrt{1 - \Omega^2}}{T \sqrt{1 - V^2}}, 
\\
\tau &=& \frac{I \Omega}{T \sqrt{(1 - V^2)(1 - \Omega^2)}}.
\end{eqnarray}
\label{eq:V-Omega-steady-state}
\end{subequations}

The reactive force and torque exerted by a circularly-polarized spin wave of angular amplitude $|\delta \mathbf n| = |\Psi| \ll 1$ are
\begin{subequations}
\begin{eqnarray}
F &=& |\Psi|^2 k_- \left[2 |r|^2 k_- + (1 - |r|^2) (k_- - k_+) \right],
\\
\tau &=& 2  |\Psi|^2 k_- (1 - |r|^2).
\end{eqnarray}
\end{subequations}
Here $k_-$ and $k_+$ are the wavenumbers of the incident ($x \to - \infty$) and transmitted ($x\to +\infty$) waves in the wall's rest frame moving at the velocity $V$ relative to the lab frame and $r$ is the reflection amplitude.  

The wavenumber and frequency of the incident wave $(\omega_-, k_-)$ in the wall frame are related to their values in the lab frame $(\omega, k)$ as follows: 
\begin{equation}
\omega_- = \frac{\omega - V k}{\sqrt{1-V^2}},
\quad
k_- = \frac{k - V \omega}{\sqrt{1-V^2}}. 
\end{equation}
In the wall frame, the frequency of the transmitted wave $\omega_+$ is redshifted from that of the incident wave $\omega_-$:
\begin{equation}
\omega_+ = \omega_- - 2 \Omega.
\end{equation}

The computation of the force and torque requires the reflection coefficient $|r|^2$ for spin waves incident upon a precessing domain wall. We have obtained it with the aid of supersymmetric quantum mechanics \cite{JPhysA.18.2917, PhysRep.251.267}. The coefficient of reflection is 
\begin{equation}
|r|^2 = \frac
		{\sinh^2{[ \frac{\pi}{2}(\tilde{k}_+ - \tilde{k}_-) ]}}
		{\sinh^2{[ \frac{\pi}{2}(\tilde{k}_+ + \tilde{k}_-) ]}},
\quad
\tilde{k}_\pm = \frac{k_\pm}{\sqrt{1-\Omega^2}}.
\label{eq:r-summary}
\end{equation}

\begin{figure}
\includegraphics[width=0.95\columnwidth]{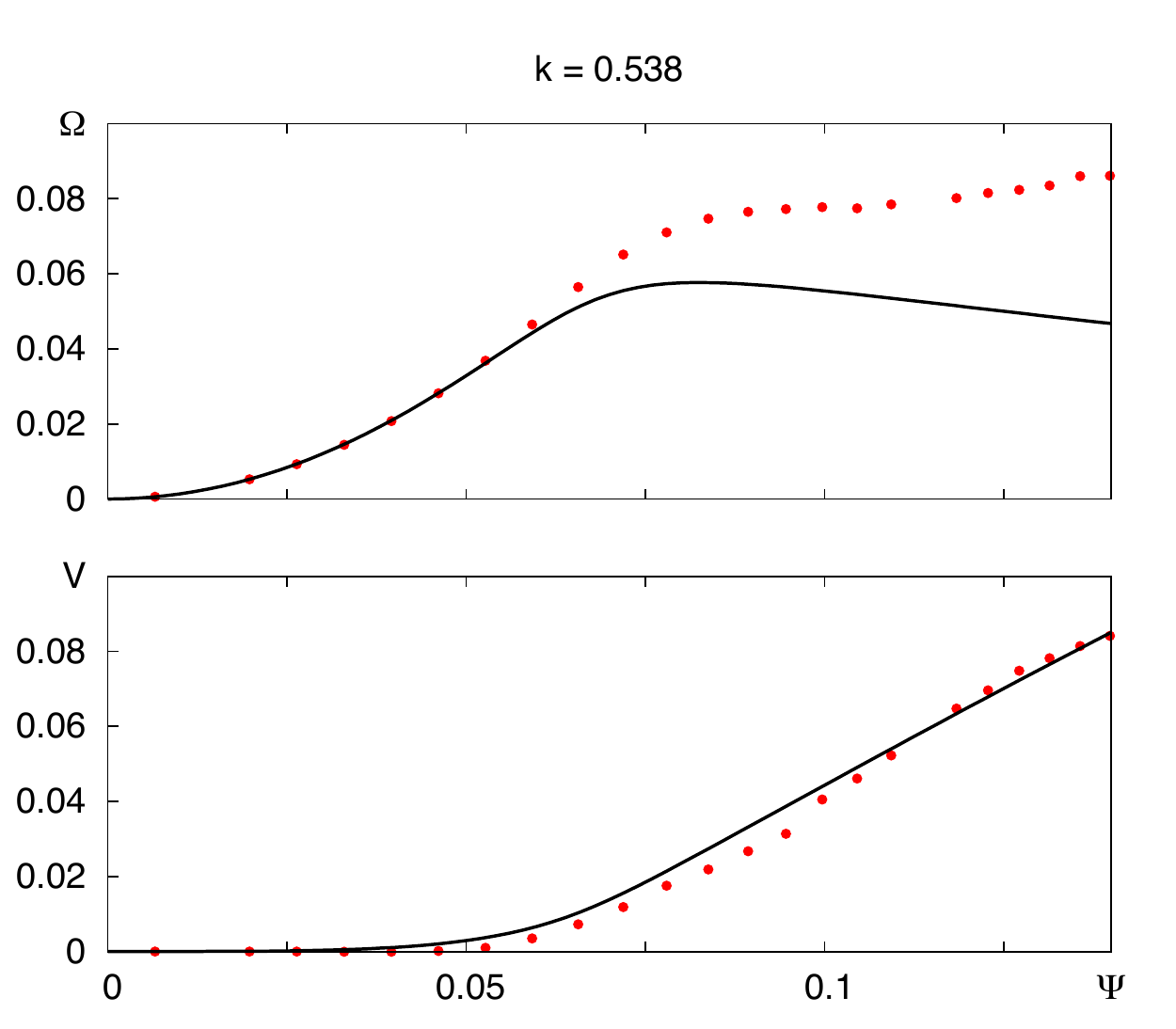}

\bigskip
\includegraphics[width=0.95\columnwidth]{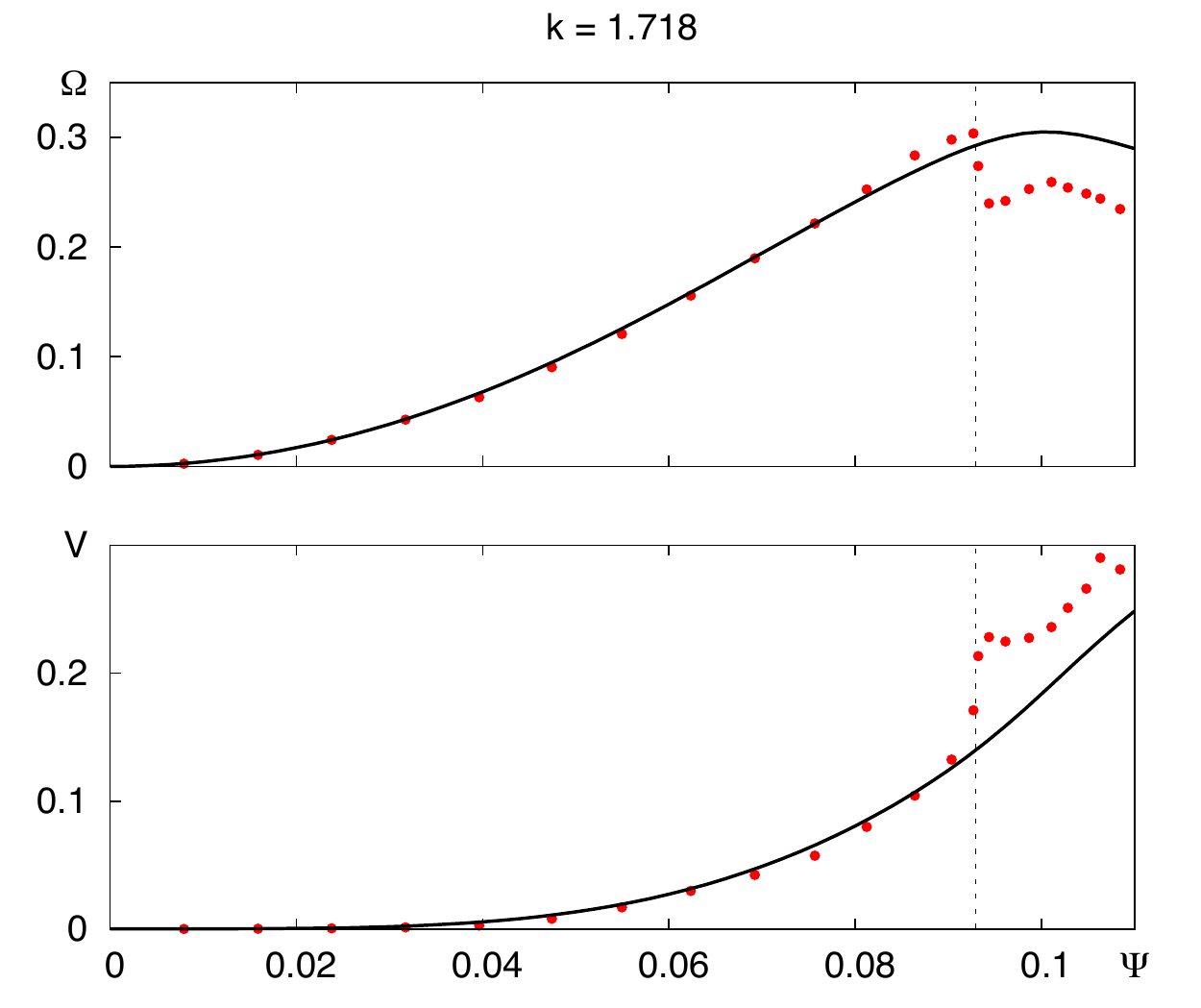}

\bigskip
\includegraphics[width=0.95\columnwidth]{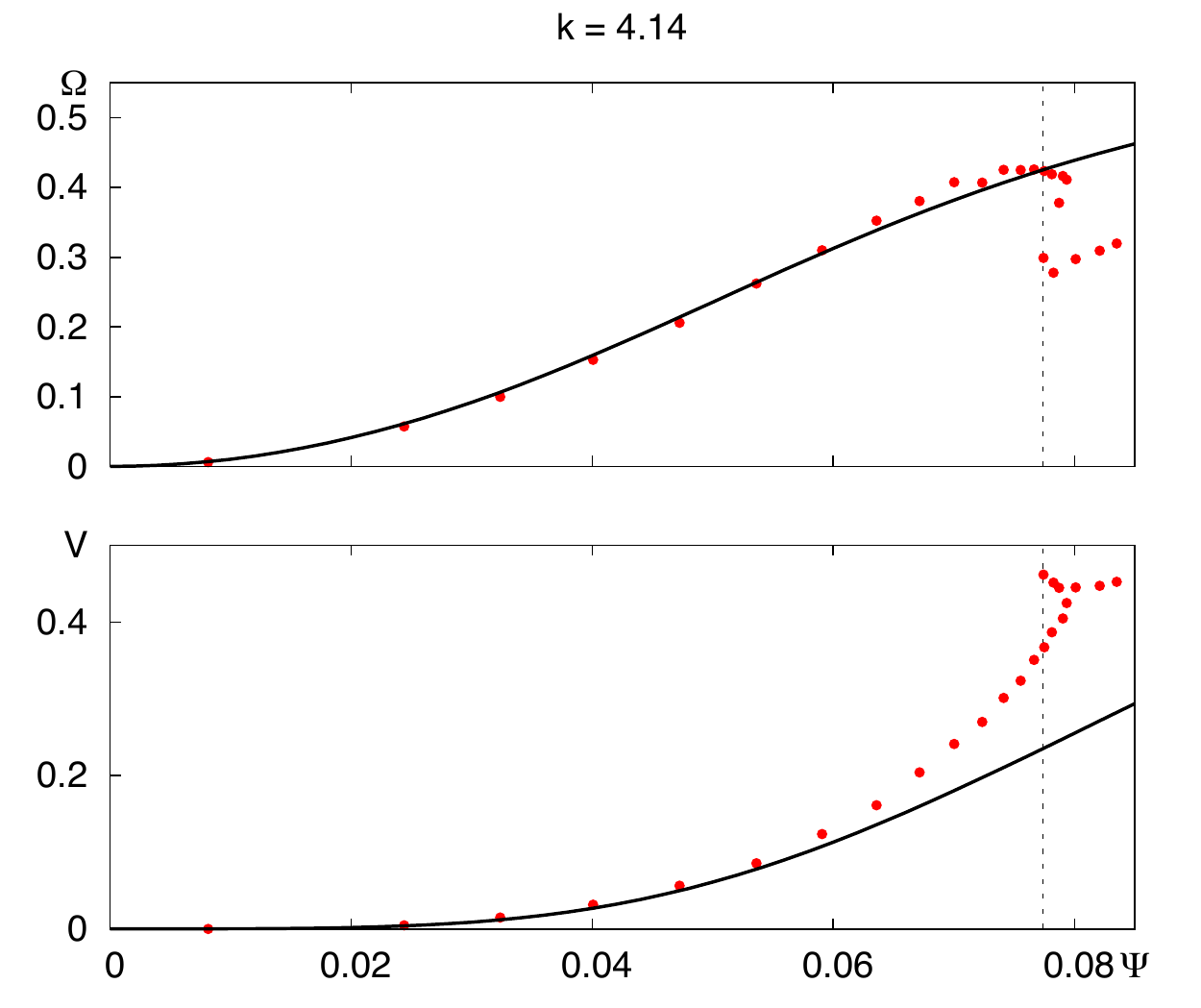}
\caption{Precession frequency and linear velocity of a domain wall as a function of the spin-wave amplitude $\Psi$ for soft, medium, and hard magnons. Solid line: theory; dots: numerical micromagnetic simulations. Vertical dotted lines: onset of a structural instability of the domain wall.}
\label{fig:comparison}
\end{figure}

Equations (\ref{eq:V-Omega-steady-state}) through (\ref{eq:r-summary}) form a closed set of algebraic equations that can be solved numerically with minimal effort. The computed dependence of the domain wall precession frequency $\Omega$ and velocity $V$ as a function of the spin-wave amplitude $\Psi$ is shown in Fig.~\ref{fig:comparison} for three different wavenumbers of the incident wave: $k = 0.538$, 1.718, and 4.14, which represent soft ($k \lesssim 1$), medium ($k \sim 1$), and hard ($k \gtrsim 1$) magnons, respectively. Generally, the primary mechanism of domain-wall propulsion switches from magnon redshift at small amplitudes to magnon reflection at large amplitudes.

The theoretical results have been tested against numerical simulations conducted with the aid of micromagnetic solver OOMMF \cite{oommf}. We have found good agreement between theory and simulations. Deviations at larger amplitudes are due to spurious reflection at the system's end (soft magnons) or due to a structural instability of the domain wall (medium and hard magnons).

\section{Ground state and domain walls}
\label{sec:qft}

\subsection{Relativistic field theory}

We consider an antiferromagnet with two sublattices whose magnetizations are described by unit-vector fields $\mathbf m_1$ and $\mathbf m_2$. Near equilibrium, the two magnetization fields are almost antiparallel, $\mathbf m_1 \approx - \mathbf m_2$, so it is convenient to describe the dynamics in terms of staggered magnetization $\mathbf n = (\mathbf m_1 - \mathbf m_2)/2$, a vector of unit length. Uniform magnetization $\mathbf m = \mathbf m_1 + \mathbf m_2$ is suppressed by the antiferromagnetic exchange interaction. It can be integrated out with the aid of its equation of motion, $\mathcal J \mathbf m = \rho \dot{\mathbf n} \times \mathbf n$, where $\mathcal J$ is the density of angular momentum on one sublattice (see Appendix~\ref{app:dynamics-general}).

The dynamics of staggered magnetization $\mathbf n$ is governed by the Lagrangian $L = \int \mathcal L \, dV$ with density \cite{PhysRevLett.50.1153}
\begin{equation}
\mathcal L = \frac{\rho |\dot{\mathbf n}|^2 - A|\mathbf n'|^2 - K_0(\hat{\mathbf z} \times \mathbf n)^2}{2}.
\label{eq:L-density-static}
\end{equation}
The Lagrangian density (\ref{eq:L-density-static}) has a ``relativistic'' form with the ``speed of light'' $s = \sqrt{A/\rho}$ \cite{PhysRevLett.50.1153}. It is invariant under Lorentz transformations
\begin{equation}
t \mapsto t' = \frac{t-vx/s^2}{\sqrt{1-v^2/s^2}},
\quad
x \mapsto x' = \frac{x-vt}{\sqrt{1-v^2/s^2}}.
\label{eq:Lorentz}
\end{equation}
This symmetry is useful for obtaining moving solutions from static ones. For example, if $\mathbf n_0(x)$ describes a static magnetic soliton minimizing the action $S = \int L \, dt$ then a Lorentz transformation yields a soliton moving at velocity $v$,
\begin{equation}
\mathbf n_v(t,x) = 
\mathbf n_0 \left(\frac{x-vt}{\sqrt{1-v^2/s^2}}\right).
\label{eq:boosted}
\end{equation}
Note that the moving soliton exhibits Lorentz contraction of its width by a factor $\sqrt{1-v^2/s^2}$.

The metric tensor and its inverse in the (1+1)-dimensional Minkowski space $(x^0,x^1) \equiv (t,x)$ are
\begin{equation}
g_{\alpha\beta} = 
	\left(\begin{array}{cc}
		s^2 & 0 \\
		0 & -1
	\end{array}\right),
\quad
g^{\alpha\beta} = 
	\left(\begin{array}{cc}
		s^{-2} & 0 \\
		0 & -1
	\end{array}\right).
\end{equation}
Some relevant physical quantities can be obtained directly from the stress-energy tensor $T^{\alpha\beta} = g^{\alpha \gamma} {T_\gamma}^\beta$, 
where 
\begin{equation}
{T_\alpha}^\beta 
	= \partial_\alpha \mathbf n \cdot \frac{\partial \mathcal L}{\partial(\partial_\beta \mathbf n)} 
		- \delta_\alpha^\beta \mathcal L.
\label{eq:def-T}
\end{equation}
Energy density, energy flux, linear momentum density, and pressure are, respectively,
\begin{eqnarray}
&& s^2 T^{00} = \frac{\rho |\dot{\mathbf n}|^2 + A|\mathbf n'|^2 + K_0(\hat{\mathbf z} \times \mathbf n)^2}{2},
\nonumber\\
&& s^2 T^{01} = - A \, \dot{\mathbf n} \cdot \mathbf n',
\quad
T^{10} = - \rho \, \dot{\mathbf n} \cdot \mathbf n', 
\label{eq:T-n}
\\
&& T^{11} = \frac{\rho |\dot{\mathbf n}|^2 + A|\mathbf n'|^2 - K_0(\hat{\mathbf z} \times \mathbf n)^2}{2}.
\nonumber
\end{eqnarray}

Axial symmetry of the problem implies conservation of angular momentum $J$. The temporal and spatial components of the associated N{\"o}ther current $j^\alpha$,
\begin{equation}
j^0 = \rho \, \hat{\mathbf z} \cdot (\mathbf n \times \dot{\mathbf n}) = - \mathcal J m_z,
\quad
j^1 = - A \, \hat{\mathbf z} \cdot (\mathbf n \times \mathbf n'),
\label{eq:j-n}
\end{equation}
are the spin density and spin current, respectively \cite{PhysRevLett.112.147204}. (The minus sign appears because magnetization and angular momentum of negatively charged electrons point in opposite directions, $\mathbf M = - \gamma \mathbf J$.) 

To simplify the expressions, we will use natural units of length, time, and energy,
\begin{equation}
\lambda_0 = \sqrt{A/K_0}, 
\quad
t_0 = \sqrt{\rho/K_0},
\quad 
\epsilon_0 = \sqrt{AK_0},
\label{eq:units}
\end{equation}
which is equivalent to setting $\rho = A = K_0 = 1$.

\subsection{Static domain wall}
\label{sec:wall-static}

Minimizing the potential energy with density 
\begin{equation}
\mathcal U = \frac{(\mathbf n')^2 + (\hat{\mathbf z} \times \mathbf n)^2}{2} 
\label{eq:L-density-U}
\end{equation} 
for boundary conditions $\mathbf{n}(\pm \infty) = \pm \hat{\mathbf z}$ yields domain-wall solutions $\mathbf n(x) = (\sin{\theta} \cos{\phi}, \sin{\theta} \sin{\phi}, \cos{\theta})$ with
\begin{equation}
\cos{\theta(x)} = \tanh{(x-X)}, 
\quad 
\phi(x) = \Phi.
\label{eq:wall-static}
\end{equation}
Collective coordinates $X$ and $\Phi$ represent the two soft modes of a domain wall, its position and azimuthal angle. The energy of a domain wall is independent of $X$ and $\Phi$, reflecting the two symmetries of the easy-axis antiferromagnet: translations and spin rotations about $\hat{\mathbf z}$.

By using the method of collective coordinates (Appendix~\ref{app:Neel}), we obtain the mass and moment of inertial of a static domain wall: 
\begin{subequations}
\begin{eqnarray}
M &=& \int |\partial \mathbf n/\partial X|^2 \, dx = 2,
\\
I &=& \int |\partial \mathbf n/\partial \Phi|^2 \, dx = 2.
\end{eqnarray}
\label{eq:M-I-collective-coords}
\end{subequations}
The energy of the domain wall can be obtained by integrating energy density (\ref{eq:T-n}): 
\[
E = \int T^{00} dx = 2.
\]

\subsection{Uniformly precessing domain wall}

Ansatz (\ref{eq:wall-static}) can be generalized for a uniformly precessing domain wall, $\Phi = \Omega t$.  To see this, we switch to a spin frame rotating at angular velocity $\bm \Omega = \Omega \hat{\mathbf z}$. If the rate of change of magnetization in the rotating frame is $\dot{\mathbf n}$ then in the lab frame it becomes $\dot{\mathbf n} + \bm \Omega \times \mathbf n$.
The Lagrangian density (\ref{eq:L-density-static}) changes to
\begin{eqnarray}
\mathcal L 
	= \frac{
		|\dot{\mathbf n} + \bm \Omega \times \mathbf n|^2 
		- |\mathbf n'|^2 
		-  (\hat{\mathbf z} \times \mathbf n)^2
	}{2}.
\end{eqnarray}
The densities of kinetic and potential energy, 
\begin{subequations}
\begin{eqnarray}
\mathcal K &=& \frac{|\dot{\mathbf n}|^2}{2} + \dot{\mathbf n} \cdot (\bm \Omega \times \mathbf n),
\label{eq:L-density-dynamic-K}
\\
\mathcal U &=& \frac{|\mathbf n'|^2 + (1 - \Omega^2)(\hat{\mathbf z} \times \mathbf n)^2}{2},
\label{eq:L-density-dynamic-U}
\end{eqnarray}
\end{subequations}
differ from the static case ($\Omega = 0$) in two ways. First, the Berry-phase term $\dot{\mathbf n} \cdot (\bm \Omega \times \mathbf n)$ encodes the Coriolis force in the rotating frame \cite{LL1}. Second, the centrifugal potential $|\bm \Omega \times \mathbf n|^2/2$ lowers the easy-axis anisotropy to $1 - \Omega^2$. The softening of anisotropy sets an upper limit to a domain wall's precession frequency, 
\begin{equation}
\Omega^2 < 1.
\label{eq:Omega-limit}
\end{equation}
At higher precession frequencies, the effective anisotropy changes sign; the switch from easy axis to easy plane destroys the domain wall. 

It is thus convenient to switch to new natural units of length, time, and energy:
\begin{equation}
\lambda_\Omega = \frac{\lambda_0}{\sqrt{1 - \Omega^2}}, 
\quad
t_\Omega = \frac{t_0}{\sqrt{1 - \Omega^2}},
\quad
\epsilon_\Omega = \epsilon_0 \sqrt{1 - \Omega^2}.
\label{eq:units-Omega}
\end{equation}
The ``speed of light'' $s = \lambda_0 / t_0 = \lambda_\Omega / t_\Omega$ is unaffected by precession. To make it clear that we use units (\ref{eq:units-Omega}), we will use variables with a tilde, e.g., 
\begin{equation}
\tilde{x} = x \sqrt{1-\Omega^2}, 
\quad
\tilde{t} = t \sqrt{1-\Omega^2},
\quad
\tilde{\omega} = \frac{\omega}{\sqrt{1-\Omega^2}}.
\end{equation}
The rescaled densities of kinetic and potential energy are thus
\begin{subequations}
\begin{eqnarray}
\tilde{\mathcal K} &=& \frac{|\partial \mathbf n/\partial \tilde{t}|^2}{2} 
	+ (\partial \mathbf n/\partial \tilde{t}) \cdot (\tilde{\bm \Omega} \times \mathbf n),
\label{eq:L-density-dynamic-K-tilde}
\\
\tilde{\mathcal U} &=& \frac{|\partial \mathbf n/\partial \tilde{x}|^2 + (\hat{\mathbf z} \times \mathbf n)^2}{2},
\label{eq:L-density-dynamic-U-tilde}
\end{eqnarray}
\end{subequations}

In the rescaled variables, potential energy density has the same form as before (\ref{eq:L-density-U}). Therefore, a domain wall static in the rotating frame has the familiar expression,
\begin{equation}
\cos{\theta} = \tanh{(\tilde{x}-\tilde{X})}, 
\quad 
\phi = \Phi.
\label{eq:wall-precessing-rotating-frame}
\end{equation}
Returning to the lab frame and natural units (\ref{eq:units}), we find
\begin{equation}
\cos{\theta} = \tanh{[(x-X)\sqrt{1-\Omega^2}]}, 
\quad 
\phi = \Phi(t) = \Omega t.
\label{eq:wall-precessing-lab-frame}
\end{equation}

The energy and angular momentum of a uniformly precessing domain wall are
\begin{eqnarray}
E &=& \int T^{00} dx = \frac{M}{\sqrt{1-\Omega^2}}, 
\label{eq:E-Omega}
\\
J &=& \int j^0 dx = \frac{I\Omega}{\sqrt{1-\Omega^2}},
\label{eq:J-Omega}
\end{eqnarray}
where $M = I = 2$ are the mass and moment of inertia of a static domain wall (\ref{eq:M-I-collective-coords}). 

\subsection{Moving domain wall}

Solutions for a domain wall moving at a constant velocity $\dot{X} = V$ can be obtained by exploiting the Lorentz symmetry, Eq.~(\ref{eq:boosted}). Viewing the energy at $V=0$ (\ref{eq:E-Omega}) as the rest mass, we readily obtain the energy and linear momentum of a moving and precessing domain wall: 
\begin{eqnarray}
E &=& \frac{M}{\sqrt{(1-\Omega^2)(1-V^2)}}, 
\label{eq:E-Omega-V}
\\
P &=& \frac{MV}{\sqrt{(1-\Omega^2)(1-V^2)}}.
\label{eq:P-Omega-V}
\end{eqnarray}
Angular momentum (\ref{eq:J-Omega}) is unchanged by the boost. The dependence of energy on linear and angular momenta is in agreement with \textcite{PhysRevLett.50.1153}: 
\begin{equation}
E^2 = 4 + J^2 + P^2.
\end{equation}

\section{Spin waves}
\label{sec:spin-waves}

\subsection{Spin waves in a ground state}

The ground states of the easy-axis antiferromagnet are 
\begin{equation}
\mathbf n_0(x) = \sigma \hat{\mathbf z},
\label{eq:Neel-states}
\end{equation}
where $\sigma = \pm 1$ is the N{\'e}el order parameter. It is convenient to use a global frame defined by three mutually orthogonal unit vectors 
\begin{equation}
\hat{\mathbf e}_1, 
\quad
\hat{\mathbf e}_2,
\quad
\hat{\mathbf e}_3 = \hat{\mathbf e}_1 \times \hat{\mathbf e}_2 = \mathbf n_0.
\label{eq:frame-ground-state}
\end{equation}

Weakly excited states can be parametrized as $\mathbf n(x) = \mathbf n_0 + \delta \mathbf n(x)$ with a small deviation $\delta \mathbf n$ orthogonal to $\mathbf n_0$. Fields 
\begin{equation}
\delta n_1 = \delta \mathbf n \cdot \hat{\mathbf e}_1,
\quad
\delta n_2 = \delta \mathbf n \cdot \hat{\mathbf e}_2
\end{equation}
describe spin waves with linear polarizations. It is convenient to introduce a complex field 
\begin{equation}
\psi = \delta \mathbf n \cdot (\hat{\mathbf e}_1 + i \hat{\mathbf e}_2).
\label{eq:psi-def}
\end{equation}
After expanding Eq.~(\ref{eq:L-density-static}) to the second order in $\psi$ we obtain the Lagrangian density for spin waves in natural units (\ref{eq:units}),
\begin{equation}
\mathcal L_\mathrm{sw} = \frac{
		|\dot{\psi}|^2
		- |\psi'|^2 
		-|\psi|^2
	}{2},
\end{equation}
which yields the equation of motion,
\begin{equation}
\ddot{\psi} = \psi'' - \psi.
\label{eq:Klein-Gordon}
\end{equation}
For a monochromatic wave with frequency $\omega$, Eq.~(\ref{eq:Klein-Gordon}) becomes an eigenproblem, $\mathcal H_0 \psi = \omega^2 \psi$, with the ``Hamiltonian'' 
\begin{equation}
\mathcal H_0 = -d^2/dx^2 + 1.
\label{eq:H-Neel}
\end{equation}
Its eigenfunctions are plane waves, 
\begin{equation}
\psi(x,t) = \Psi \, e^{-i \omega t + i k x},
\label{eq:plane-wave}
\end{equation}
with a ``relativistic'' dispersion
\begin{equation}
\omega^2 = 1 + k^2.
\label{eq:dispersion}
\end{equation}

\begin{figure}
\includegraphics[width=0.95\columnwidth]{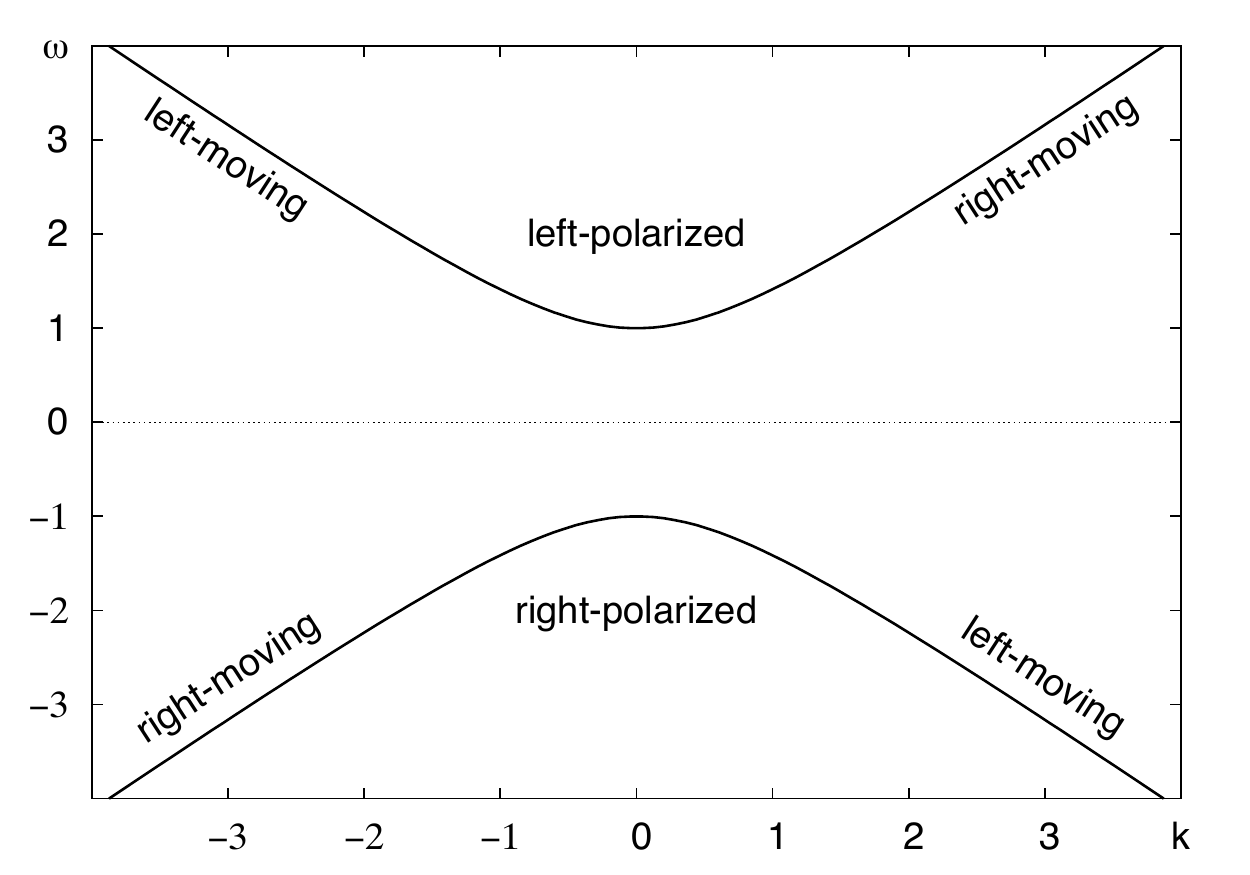}
\caption{Dispersion $\omega(k)$ of circularly polarized spin waves (\ref{eq:plane-wave}) in a uniaxial antiferromagnet in natural units of length and time (\ref{eq:units}).}
\label{fig:dispersion}
\end{figure}

Equation (\ref{eq:plane-wave}) describes a circularly polarized wave with the amplitude $|\delta \mathbf n| = \Psi$. If $\omega < 0$, $\delta \mathbf n$ precesses from $\hat{\mathbf e}_1$ to $\hat{\mathbf e}_2$. We will call such waves right-circularly polarized. Waves with $\omega > 0$ precess from $\hat{\mathbf e}_2$ to $\hat{\mathbf e}_1$ and will be called left-circularly polarized. The group velocity, 
\begin{equation}
v = d\omega/dk = k/\omega,
\end{equation}
determines the direction of propagation: waves with $k/\omega > 0$ are right-moving, waves with $k/\omega < 0$ are left-moving, Fig.~\ref{fig:dispersion}. 

A circularly polarized spin wave (\ref{eq:plane-wave}) in the background of a N{\'e}el ground state (\ref{eq:Neel-states}) has spin density and current 
\begin{equation}
j^0 = - \sigma \omega |\Psi|^2,
\quad
j^1 = - \sigma k |\Psi|^2.
\label{eq:j-uniform}
\end{equation}
The density of linear momentum and pressure are
\begin{equation}
T^{10} = \omega k |\Psi|^2, 
\quad
T^{11} = k^2 |\Psi|^2.
\label{eq:T-uniform}
\end{equation}
Note that spin density and current (\ref{eq:j-uniform}) depend on the N{\'e}el ground state (\ref{eq:Neel-states}) through the order parameter $\sigma = \pm 1$, whereas linear momentum density and pressure (\ref{eq:T-uniform}) do not.

\subsection{Spin waves on a static domain wall}

A static domain wall is a local minimum of potential energy with $\mathbf n_0(x)$ interpolating between $-\hat{\mathbf z}$ and $+\hat{\mathbf z}$. The N{\'e}el order parameter $\sigma$ interpolates between $-1$ and $+1$. In spherical coordinates, $\mathbf n_0 = (\sin{\theta} \cos{\phi}, \, \sin{\theta} \sin{\phi}, \, \cos{\theta})$,  
\begin{equation}
\cos{\theta} = \tanh{(x-X)}, 
\ 
\sin{\theta} = \mathop{\mathrm{sech}}{(x-X)}, 
\ 
\phi = \Phi, 
\label{eq:domain-wall-static}
\end{equation}
where $X$ and $\Phi$ are arbitrary position and angle, Fig.~\ref{fig:domain-wall}. We will set $X = 0 $ and $\Phi = 0$ to obtain spin-wave solutions.

Next we consider a small-amplitude spin wave $\delta \mathbf n$ in the background of a static domain wall $\mathbf n_0$ (\ref{eq:domain-wall-static}). Again, $\delta \mathbf n$ is transverse to $\mathbf n_0$ and can be expressed in terms of the complex field $\psi$ (\ref{eq:psi-def}), where unit vectors $\hat{\mathbf e}_1$ and $\hat{\mathbf e}_2$ are orthogonal to $\mathbf n_0$. A convenient local frame is 
\begin{equation}
\hat{\mathbf e}_1 = \frac{\partial \mathbf n_0}{\partial \theta}, 
\quad
\hat{\mathbf e}_2 = \frac{\partial \mathbf n_0}{\sin{\theta} \, \partial \phi},
\quad
\hat{\mathbf e}_3 = \hat{\mathbf e}_1 \times \hat{\mathbf e}_2 = \mathbf n_0, 
\label{eq:local-frame}
\end{equation}
with $\mathbf n_0$ given by Eq.~(\ref{eq:domain-wall-static}). An expansion of the Lagrangian in powers of $\psi$ yields the following Lagrangian density for spin waves: 
\begin{equation}
\mathcal L_\mathrm{sw} = \frac{
		|\dot{\psi}|^2
		- |\psi'|^2 
		-[1 - 2 \mathop{\mathrm{sech}^2}(x)] \, |\psi|^2
	}{2}.
\label{eq:L-sw-static}
\end{equation}

\begin{figure}
\includegraphics[width=0.95\columnwidth]{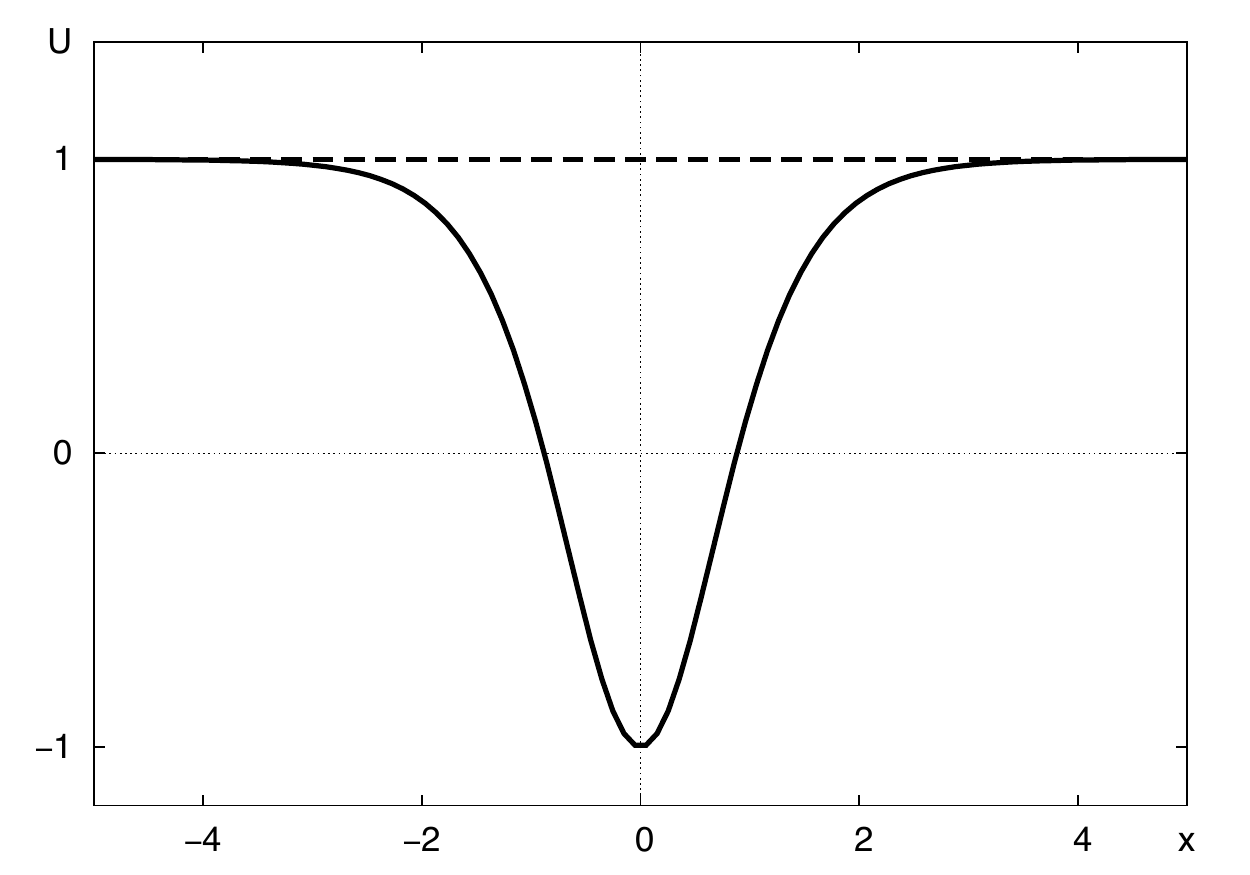}
\caption{Solid line: the P{\"o}schl-Teller potential (\ref{eq:H-domain-wall}). Dashed line: its SUSY partner (\ref{eq:H-Neel}).}
\label{fig:susy1}
\end{figure}

The equation of motion for spin waves are
\begin{equation}
\ddot{\psi} 
	= \psi''  - \left[ 1 - 2 \, \mathop{\mathrm{sech}^2} (x) \right]\psi, 
\label{eq:Schroedinger-static-background}
\end{equation}
A monochromatic wave $\psi(x,t) = \psi(x) e^{-i\omega t}$ satisfies the ``Schr{\"o}dinger equation'' $\mathcal H_1 \psi = \omega^2 \psi$ with the ``Hamiltonian''
\begin{equation}
\mathcal H_1 = -d^2/dx^2 + 1 - 2 \mathop{\mathrm{sech}^2}{x}.
\label{eq:H-domain-wall}
\end{equation}
Comparison to its counterpart in the uniform ground state (\ref{eq:H-Neel}) shows that the presence of the domain wall creates a potential well for spin waves, $U_\mathrm{dw} = - 2 \mathop{\mathrm{sech}^2}{x}$, Fig.~\ref{fig:susy1}. This potential, named after \textcite{ZPhys.83.143}, has a remarkable property: waves pass through it without reflection. 

\subsubsection{Supersymmetric solution}

The exact solvability of the P{\"o}schl-Teller problem and the lack of reflection can be traced to a special relation---known as supersymmetry (SUSY) \cite{JPhysA.18.2917, PhysRep.251.267}---between its ``Hamiltonian'' (\ref{eq:H-domain-wall}) and that of a free particle (\ref{eq:H-Neel}). Both ``Hamiltonians'' can be factorized in terms of the same operators $a$ and $a^\dagger$:
\begin{eqnarray}
\mathcal H_0 = a a^\dagger,
&\quad&
\mathcal H_1 = a^\dagger a,
\\
a = d/dx + \tanh{x},
&\quad&
a^\dagger = - d/dx + \tanh{x}.
\end{eqnarray}
Eigenstates of the free ``Hamiltonian'' $\mathcal H_0$ are plane waves, $\psi_0(x) = \Psi e^{i k x}$, with $\omega^2 = 1 + k^2$. Eigenstates of $\mathcal H_1$ can be obtained from those of $\mathcal H_0$: 
$\psi_1(x) = a^\dagger \psi_0(x)$. Furthermore, SUSY partners $\psi_0$ and $\psi_1$ have the same eigenfrequency:
\[
\mathcal H_1 \psi_1 = a^\dagger a (a^\dagger \psi_0) = a^\dagger (a a^\dagger \psi_0)
	= a^\dagger (\omega^2 \psi_0) = \omega^2 \psi_1.
\] 
This yields a spin wave in the background of a static domain wall \cite{Yan2011, PhysRevLett.112.147204}: 
\begin{equation}
\psi(x,t) = \Psi \, \frac{\tanh{x} - i k}{-1 - i k} e^{- i \omega t + i k x},
\quad
\omega^2 = 1 + k^2.
\label{eq:spin-wave-static-background}
\end{equation}
Eq.~(\ref{eq:spin-wave-static-background}) describes a circularly polarized spin wave with amplitude $\Psi$ at $x = -\infty$ and $\Psi e^{i \delta}$ at $x=+\infty$. It passes through the domain wall without reflection and picks up a phase shift $\delta = 2 \arctan{(1/k)}$. 

\subsubsection{Reactive force and torque}
\label{eq:sec-static-influence}

To study the effect of the spin wave on the domain wall, let us consider a left-circularly polarized ($\omega>0$) spin wave incoming from the left ($k>0$). After recalling that polarization is defined in the local frame (\ref{eq:frame-ground-state}) tied to the staggered magnetization $\hat{\mathbf n}_0$ (\ref{eq:Neel-states}) and that $\hat{\mathbf n}_0$ is reversed by a domain wall, we find that the direction of precession in the global frame is reversed, too: the wave precesses from $\hat{\mathbf x}$ to $\hat{\mathbf y}$ on the left of the domain wall (where $\hat{\mathbf n}_0 = - \hat{\mathbf z}$) and from $\hat{\mathbf y}$ to $\hat{\mathbf x}$ on the right (where $\hat{\mathbf n}_0 = + \hat{\mathbf z}$). Spin density and current (\ref{eq:j-uniform}) are positive on the left (where $\sigma = -1$) and negative on the right (where $\sigma = +1$). In contrast, linear momentum density (\ref{eq:T-uniform}) is positive on both sides of the domain wall. 

In the language of quantum mechanics, the spin wave contains magnons with angular momentum $J = +\hbar$ and linear momentum $P = + \hbar k$ on the left of the domain wall; $J = -\hbar$ and $P = + \hbar k$ on the right. Because all magnons pass through the domain wall and their linear momenta remain unchanged, the spin wave exerts no force on the domain wall. The reversal of their angular momenta means that each magnon deposits spin $+ 2\hbar$ on the domain wall, thereby exerting a positive torque on it. The torque can be computed as the net spin current into the domain wall \cite{PhysRevLett.112.147204},
\begin{equation}
\tau = j^1(-\infty) - j^1(+\infty) = 2 k |\Psi|^2. 
\label{eq:tau-static-wall}
\end{equation}
Under zero net force and finite torque, the domain wall retains zero linear momentum and thus zero linear velocity but acquires a finite angular momentum and angular velocity. 

We can estimate the angular frequency of precession from the balance of the spin-wave and viscous torques, $\tau - D_{\Phi\Phi} \Omega = 0$, in a steady state (see Appendix~\ref{app:Neel} for details). The dissipation coefficient is related to the moment of inertia by Eq.~(\ref{eq:D-M-relation}): $D_{\Phi\Phi} = M_{\Phi\Phi}/T \equiv I/T$. Hence 
\begin{equation}
\Omega = k T |\Psi|^2. 
\end{equation}
This result is confirmed by a more detailed analysis in Sec.~\ref{sec:analysis-soft}, see Eq.~(\ref{eq:Omega-V-soft-redshift}).

\subsection{Spin waves on a precessing domain wall}

Expanding the Lagrangian for small fluctuations in the vicinity of the domain-wall solution in the rotating frame yields the following to the second order in $\psi$:
\begin{eqnarray}
\tilde{\mathcal L}_\mathrm{sw} &=& \frac{
		|\partial \psi / \partial \tilde{t}|^2
		- |\partial \psi / \partial \tilde{x}|^2 
		- (1 - 2 \mathop{\mathrm{sech}^2}\tilde{x}) \, |\psi|^2
	}{2}
\nonumber
\\
&& - i \tilde{\Omega} \tanh{\tilde{x}} \, \psi^* \partial \psi / \partial \tilde{t}.
\label{eq:L-sw-dynamic}
\end{eqnarray}
The factor $\tilde{\Omega} \tanh{\tilde{x}}$ in front of the Berry-phase term is the projection of the frame angular velocity $\tilde{\bm \Omega} = \tilde{\Omega} \hat{\mathbf z}$ onto the local precession axis $\hat{\mathbf e}_3 = \mathbf n_0$. The spin-wave equation for a monochromatic wave $\psi(\tilde{x},\tilde{t}) = \psi(x) \exp{(-i \tilde{\omega} \tilde{t})}$ in the rotating frame is 
\begin{eqnarray}
\tilde{\omega}^2 \psi 
	&=&  - \partial^2 \psi / \partial \tilde{x}^2  
		+ ( 1 - 2 \, \mathop{\mathrm{sech}^2}{\tilde{x}} )\psi 
\nonumber\\
	&& + 2 \tilde{\omega} \tilde{\Omega} \tanh{\tilde{x}} \, \psi. 
\label{eq:Schroedinger-rotating-background}
\end{eqnarray}

\subsubsection{Supersymmetric solution}

\begin{figure}
\includegraphics[width=0.95\columnwidth]{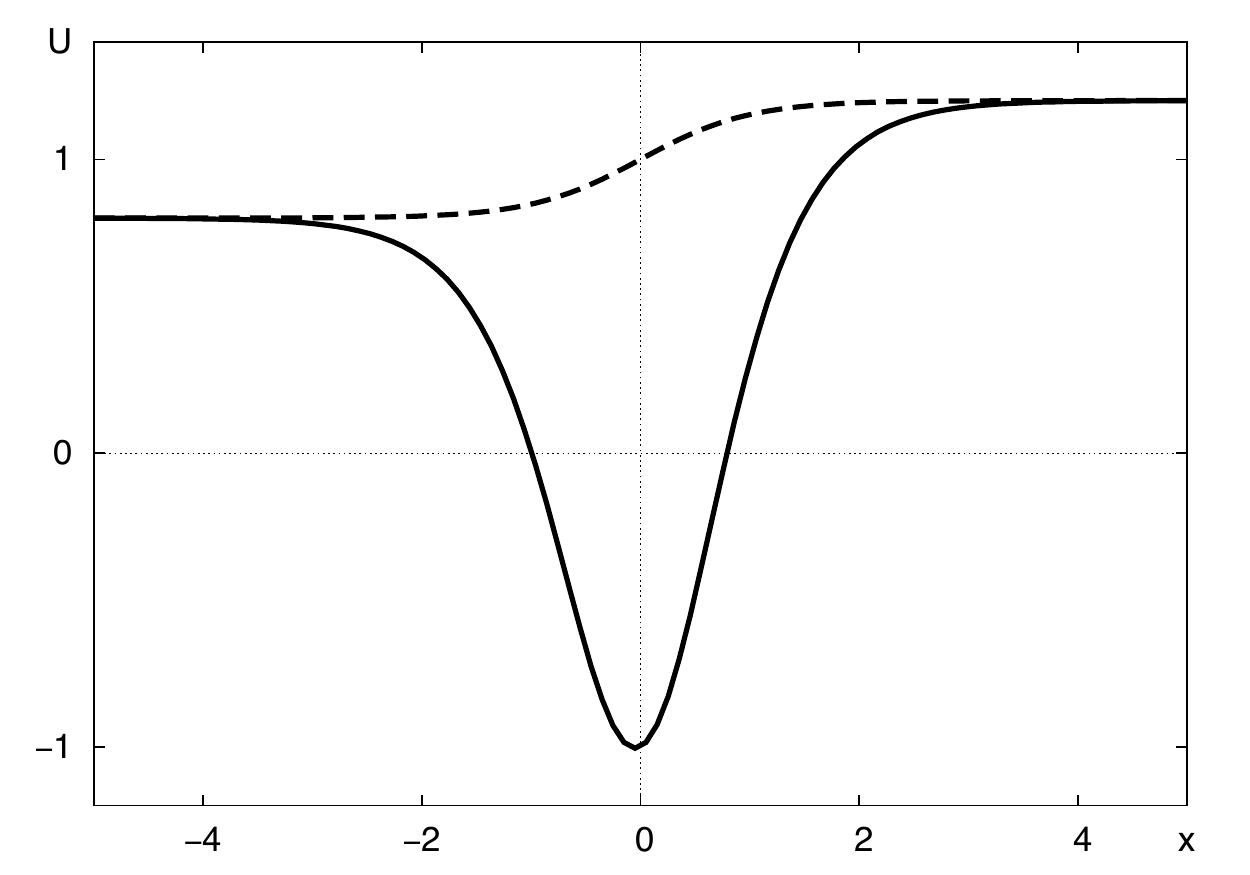}
\caption{Solid line: the spin-wave potential on a rotating domain wall (\ref{eq:H-SUSY-rotating}). Dashed line: its SUSY partner (\ref{eq:H-SUSY-rotating-partner}).}
\label{fig:susy2}
\end{figure}

The spin-wave ``Hamiltonian'' is
\begin{equation}
\mathcal H_1 = - d^2/d\tilde{x}^2 + 1 - 2 \mathop{\mathrm{sech}^2}{\tilde{x}} 
	+ 2 \tilde{\omega} \tilde{\Omega} \tanh{\tilde{x}}. 
\label{eq:H-SUSY-rotating}
\end{equation}
As in the static case, we express it in terms of raising and lowering operators,
\begin{equation}
\mathcal H_1 = a^\dagger a - (\tilde{\omega} \tilde{\Omega})^2,
\end{equation}
where
\begin{equation}
a= d/d\tilde{x} + \tanh{\tilde{x}} + \tilde{\omega} \tilde{\Omega}, 
\ 
a^\dagger = -d/d\tilde{x} + \tanh{\tilde{x}} + \tilde{\omega} \tilde{\Omega}.
\end{equation}
Its supersymmetric partner ``Hamiltonian'' is 
\begin{equation}
\mathcal H_0 = a a^\dagger - (\tilde{\omega} \tilde{\Omega})^2 
	= - d^2/d\tilde{x}^2 + 1 + 2 \tilde{\omega} \tilde{\Omega} \tanh{\tilde{x}}.
\label{eq:H-SUSY-rotating-partner}
\end{equation}
Potential energies of $\mathcal H_0$ and $\mathcal H_1$ are shown in Fig.~\ref{fig:susy2}. 

Whereas in the static situation the SUSY partner $\mathcal H_0$ was the Hamiltonian of a free particle (\ref{eq:H-Neel}), the precession of the domain wall creates a smoothed step potential $2 \tilde{\omega} \tilde{\Omega} \tanh{\tilde{x}}$. This has two important ramifications. First, the wavenumbers on the left and on the right of the domain wall ($\tilde{k}_\pm$ for $\tilde{x} \to \pm \infty$) are no longer the same: 
\begin{equation}
\tilde{k}_{\pm}^2 = \tilde{\omega}^2 \mp 2 \tilde{\omega} \tilde{\Omega} - 1.
\label{eq:k-omega-rotating-frame}
\end{equation}
Second, because $\mathcal H_0$ is no longer reflection-free, $\mathcal H_1$ also exhibits reflection of spin waves. Furthermore, as we show in Appendix \ref{app:r-t}, their reflection and transmission amplitudes $r$ and $t$ have the same absolute values. The reflection amplitude squared is
\begin{equation}
|r|^2 = \frac
		{\sinh^2{\left[ \frac{\pi}{2}(\tilde{k}_+ - \tilde{k}_-) \right]}}
		{\sinh^2{\left[ \frac{\pi}{2}(\tilde{k}_+ + \tilde{k}_-) \right]}}.
\label{eq:r-exact}
\end{equation}
The transmission amplitude can be found from it with the aid of the identity
\begin{equation}
|r|^2 \tilde{k}_- + |t|^2 \tilde{k}_+ = \tilde{k}_-.
\label{eq:r-t-identity}
\end{equation}
Note that the wavevectors squared (\ref{eq:k-omega-rotating-frame}) cannot be negative because of the upper limit (\ref{eq:Omega-limit}) on a domain wall's precession frequency.

\subsubsection{Return to the lab frame}

To use the results in elaborating a theory for propulsion, we need to return from the rotating frame to the lab frame. First, we return to the natural units of the static frame (\ref{eq:units}) to obtain 
\begin{equation}
k_{\pm}^2 = (\omega \mp \Omega)^2 - 1.
\label{eq:k-omega-2}
\end{equation}
As we mentioned at the beginning of Sec.~\ref{eq:sec-static-influence}, the spin wave is precessing in opposite directions on the two sides of the domain wall: from $\hat{\mathbf x}$ to $\hat{\mathbf y}$ on the left of the domain wall and from $\hat{\mathbf y}$ to $\hat{\mathbf x}$ on the right. The domain wall rotates from $\hat{\mathbf x}$ to $\hat{\mathbf y}$. Upon returning to the lab frame, we find that the wave is precessing faster on the left, with the frequency $\omega_- = \omega + \Omega$ and slower on the right, with the frequency $\omega_+ = \omega - \Omega$. We can check that $k_{\pm}^2 = \omega_\pm^2 - 1$ (\ref{eq:k-omega-2}) agrees with the spin-wave dispersion (\ref{eq:dispersion}). In the lab frame, the incoming wave of frequency $\omega_-$ is transmitted through the rotating domain wall with a redshift: 
\begin{equation}
\omega_+ = \omega_- - 2 \Omega.
\label{eq:redshift}
\end{equation}
The frequency of the reflected wave is the same as that of the incoming wave---in the frame where the domain wall has no translational motion. A nonzero velocity of the domain wall would create a Doppler shift for the reflected wave. 

\section{Forces and torques}
\label{sec:forces-torques}

\subsection{Reactive force and torque}

The reactive force and torque exerted by spin waves on a domain wall can be computed from conservation laws of linear and angular momenta expressed by zero divergence of the energy-momentum tensor and spin current, $\partial_\beta T^{\alpha\beta} = 0$, $\partial_\alpha j^\alpha = 0$. Integrating these identities over $x$ yields 
\begin{subequations}
\begin{eqnarray}
\dot{P} &=& T^{11}(-\infty) - T^{11}(+\infty) \equiv F, 
\\
\dot{J} &=& j^1(-\infty) - j^1(+\infty) \equiv \tau.
\end{eqnarray}
\label{eq:reactive-F-tau}
\end{subequations}
The right-hand sides of Eqs.~(\ref{eq:reactive-F-tau}) are the reactive force and torque given by the pressure difference on the two sides of the domain wall and by the net spin current flowing into it, respectively. 

Equations (\ref{eq:reactive-F-tau}) are most conveniently applied in an inertial frame moving at the instantaneous velocity the domain wall. In that frame, the linear and angular momenta of the spin wave remain unchanged and so $\dot{P}$ and $\dot{J}$ on the left-hand sides are those of the domain wall alone.  In other frames, the motion of the domain wall alters the configuration of the spin wave; $\dot{P}$ and $\dot{J}$ incorporate changes in linear and angular momenta of the spin wave \cite{PhysRevLett.112.147204}. 

Using the results from the previous section, we obtain the reactive force and torque,
\begin{subequations}
\begin{eqnarray}
F &=& |\Psi|^2 k_- \left[2 |r|^2 k_- + (1 - |r|^2) (k_- - k_+) \right], \label{eq:F}
\\
\tau &=& 2  |\Psi|^2 k_- (1 - |r|^2).
\end{eqnarray}
\label{eq:F-tau}
\end{subequations}
Here $|\Psi|^2 k_-$ is the incoming spin current; $|r|^2$ is the probability of reflection for a magnon, while $1-|r|^2$ the probability of transmission; $2k_-$ is the momentum lost by a reflected magnon, and $k_- - k_+$ is the same for a transmitted one. We may thus ascribe the two terms in Eq.~(\ref{eq:F-tau}a) to reflection and redshift:
\begin{eqnarray}
F_{\mathrm{reflection}} &=& 2 |\Psi|^2 |r|^2 k_-^2,
\nonumber\\
F_{\mathrm{redshift}} &=& |\Psi|^2 (1 - |r|^2) k_- (k_- - k_+).
\label{eq:F-parts}
\end{eqnarray}
The torque is entirely due to transmitted magnons: reflected ones keep their angular momentum.

The relative contributions of reflection and redshift to the reactive force are 
\begin{equation}
\frac{F_\mathrm{reflection}}{F_\mathrm{redshift}} 
	= \frac{|r|^2}{1-|r|^2} \, \frac{2 k_-}{k_- - k_+}.
\end{equation}
For hard magnons, $k_\pm \gg 1$, the dominant contribution comes from redshift. The coefficient of reflection (\ref{eq:r-exact}) is exponentially suppressed, $|r|^2 \sim e^{- 2 \pi k_+} \ll 1$, and
\begin{equation}
\frac{F_\mathrm{reflection}}{F_\mathrm{redshift}} 
	\sim \frac{2 k_- e^{- 2 \pi k_+}}{k_- - k_+}
	\ll 1.
\quad 
\mbox{(hard magnons)} 
\label{eq:r-hard}
\end{equation} 
Reflection can dominate if both magnons are soft, $k_{\pm} \ll 1$. In this ``non-relativistic'' limit, the ratio 
\begin{equation}
\frac{F_\mathrm{reflection}}{F_\mathrm{redshift}} 
	\sim \frac{k_- - k_+}{2 k_+} 
\quad
\mbox{(soft magnons)}
\label{eq:r-soft}
\end{equation}
is also small unless transmitted magnons experience a substantial redshift, $k_+ \ll k_-$. 

From Eq.~(\ref{eq:r-hard}) one might infer that reflection is also important if incoming magnons are hard, $k_- \gg 1$, and transmitted magnons are soft, $k_+ \ll 1$. However, on account of Eq.~(\ref{eq:redshift}) this would require fast precession of the domain wall, 
\begin{equation}
\Omega = (\omega_- - \omega_+)/2 \sim (k_- - 1)/2 \gg 1. 
\end{equation}
A domain wall precessing faster than $\Omega = 1$ disintegrates as in Eq.~(\ref{eq:Omega-limit}) because a strong centrifugal force in the rotating frame turns anisotropy from easy-axis to easy-plane. For this reason, a situation where incoming magnons are hard and transmitted magnons are soft never occurs (in the reference frame where the domain wall has zero linear velocity). 

Lastly, we relate the frequency and  wavenumber of the incoming magnons measured in the lab frame $(\omega, k)$ to those in the moving frame $(\omega_-, k_-)$ by the Lorentz transformation:
\begin{equation}
\omega_- = \frac{\omega - V k}{\sqrt{1-V^2}}, 
\quad
k_- = \frac{k - V \omega}{\sqrt{1-V^2}}.
\label{eq:Lorentz-transform}
\end{equation}

\subsection{Viscous force and torque}

The inclusion of dissipation violates conservation of linear and angular momenta. Eqs.~(\ref{eq:reactive-F-tau}) acquire additional contributions in the form of viscous force $F^v$ and torque $\tau^v$. Our task here is to compute them in the inertial frame where the domain wall is momentarily at rest, as in the previous section. 

In the presence of dissipation, the Euler-Lagrange equations are modified by a viscous friction term \cite{LL1},
\begin{equation}
\partial_\alpha \frac{\partial \mathcal L}{\partial (\partial_\alpha \mathbf n)}
- \frac{\partial \mathcal L}{\partial \mathbf n}  + \frac{\partial \mathcal R}{\partial \dot{\mathbf n}} 
= 0,
\label{eq:Euler-Lagrange-Rayleigh-lab-frame}
\end{equation}
where $\mathcal R = |\dot{\mathbf n}|^2/(2T)$ is the density of Rayleigh's dissipation function (\ref{eq:R-antiferro}) and $T$ is the relaxation time~(\ref{eq:relaxation-time}). Eq.~(\ref{eq:Euler-Lagrange-Rayleigh-lab-frame}) is valid in the frame of the antiferromagnet and is not Lorentz-invariant since there is now a preferred reference frame. However, we can give it a Lorentz-invariant form if we recast the viscous term as follows:
\[
\frac{\partial \mathcal R}{\partial \dot{\mathbf n}} 
	= \frac{\dot{\mathbf n}}{T} 
	= \frac{u^\alpha \partial_\alpha \mathbf n}{T},
\]
where $u = (1, 0)$ is the 2-velocity of the antiferromagnet in the lab frame. Now the Euler-Lagrange equation with dissipation has a Lorentz-invariant form:  
\begin{equation}
\partial_\alpha \frac{\partial \mathcal L}{\partial (\partial_\alpha \mathbf n)}
- \frac{\partial \mathcal L}{\partial \mathbf n}  + \frac{u^\alpha \partial_\alpha \mathbf n}{T}
= 0.
\label{eq:Euler-Lagrange-Rayleigh-generic}
\end{equation}
In a frame moving with velocity $v$, the modified Euler-Lagrange equation (\ref{eq:Euler-Lagrange-Rayleigh-generic}) remains the same, while the 2-velocity of the antiferromagnet becomes
\begin{equation}
u = \left(\frac{1}{\sqrt{1-v^2}}, \, -\frac{v}{\sqrt{1-v^2}}\right).
\end{equation}

Viscous losses associated with Gilbert damping break conservation laws of linear and angular momenta. This violation is manifested in non-zero divergence of the energy-momentum tensor and current: 
\begin{subequations}
\begin{eqnarray}
\partial_\beta T^{\alpha\beta} &=& - u^\beta \partial^\alpha \mathbf n \cdot \partial_\beta \mathbf n/T,
\\
\partial_\alpha j^\alpha &=& - u^\alpha \, \hat{\mathbf z} \cdot (\mathbf n \times \partial_\alpha \mathbf n)/T.
\end{eqnarray}
\end{subequations}
Integration over $x$ yields the viscous force and torque in a moving frame: 
\begin{subequations}
\begin{eqnarray}
F^v &=& D_{1\alpha} u^\alpha,
\\
\tau^v &=& - N_\alpha u^\alpha,
\end{eqnarray}
\end{subequations}
where 
\begin{eqnarray}
D_{\alpha\beta} &=& T^{-1} \int \partial_\alpha \mathbf n \cdot \partial_\beta \mathbf n \, dx,
\\
N_{\alpha} &=& T^{-1} \int \hat{\mathbf z} \cdot (\mathbf n \times \partial_\alpha \mathbf n) \, dx.
\end{eqnarray}

In the lab frame, where $u = (1,0)$, we only need components $D_{10}$ and $N_0$, which are directly related to linear and angular momenta: $D_{10} = -P/T$, $N_0 = - J/T$. We thus obtain equations of motion in the lab frame:
\begin{subequations}
\begin{eqnarray}
\dot{P} &=& F - P/T, 
\\ 
\dot{J} &=& \tau - F/T,
\label{eq:eom-P-J-lab}
\end{eqnarray}
\end{subequations}
which yields the viscous force $F^v = - P/T$ and torque $F^v = - J/T$ in the lab frame.

In the frame moving at the velocity of the domain wall $V$, the wall has zero linear velocity, so that $D_{10} = P/T = 0$, $D_{11} = M \sqrt{1- \Omega^2}$, and $N_1 = 0$. We then obtain 
\begin{subequations}
\begin{eqnarray}
\dot{P} &=& F - \frac{MV \sqrt{1-\Omega^2}}{T\sqrt{1-V^2}}, 
\\ 
\dot{J} &=& \tau - \frac{J}{T\sqrt{1-V^2}}.
\label{eq:eom-P-J-wall}
\end{eqnarray}
\end{subequations}
In a steady state, 
\begin{subequations}
\begin{eqnarray}
F &=& \frac{M V \sqrt{1 - \Omega^2}}{T \sqrt{1 - V^2}},
\\
\tau &=& \frac{I \Omega}{T \sqrt{(1 - V^2)(1 - \Omega^2)}},
\end{eqnarray}
\label{eq:eom-V-Omega-steady}
\end{subequations}
where we have used the expression for angular momentum of the domain wall (\ref{eq:J-Omega}).

\section{Analysis of the dynamics}
\label{sec:analysis}

Equations~(\ref{eq:eom-V-Omega-steady}) determine the steady-state dynamics of a domain wall driven by a circularly polarized spin wave. Together with the expressions for the reactive force and torque (\ref{eq:F-tau}) and the coefficient of reflection (\ref{eq:r-exact}), these equations constitute the core formal results of our theory. These equations can be readily solved by numerical means. In certain limits the expressions simplify and we can make further progress analytically. This is the case in the regimes where the reactive force is dominated either by reflection or by redshift. 

In this section we discuss the dynamics of a domain wall as a function of the spin-wave amplitude $\Psi$ at a fixed wavenumber $k$, focusing on the limits of soft and hard spin waves. In both limits, the primary mechanism of propulsion switches from redshift at small amplitudes to reflection at large amplitudes. For simplicity, we work in the limit where the linear and angular velocities of the domain wall are small in natural units (\ref{eq:units}), $V \ll 1$ and $\Omega \ll 1$. We discuss separately the cases of soft ($k \ll 1$) and hard ($k \gg 1$) spin waves. 

\subsection{Soft magnons}
\label{sec:analysis-soft}

\begin{figure}
\includegraphics[width=0.95\columnwidth]{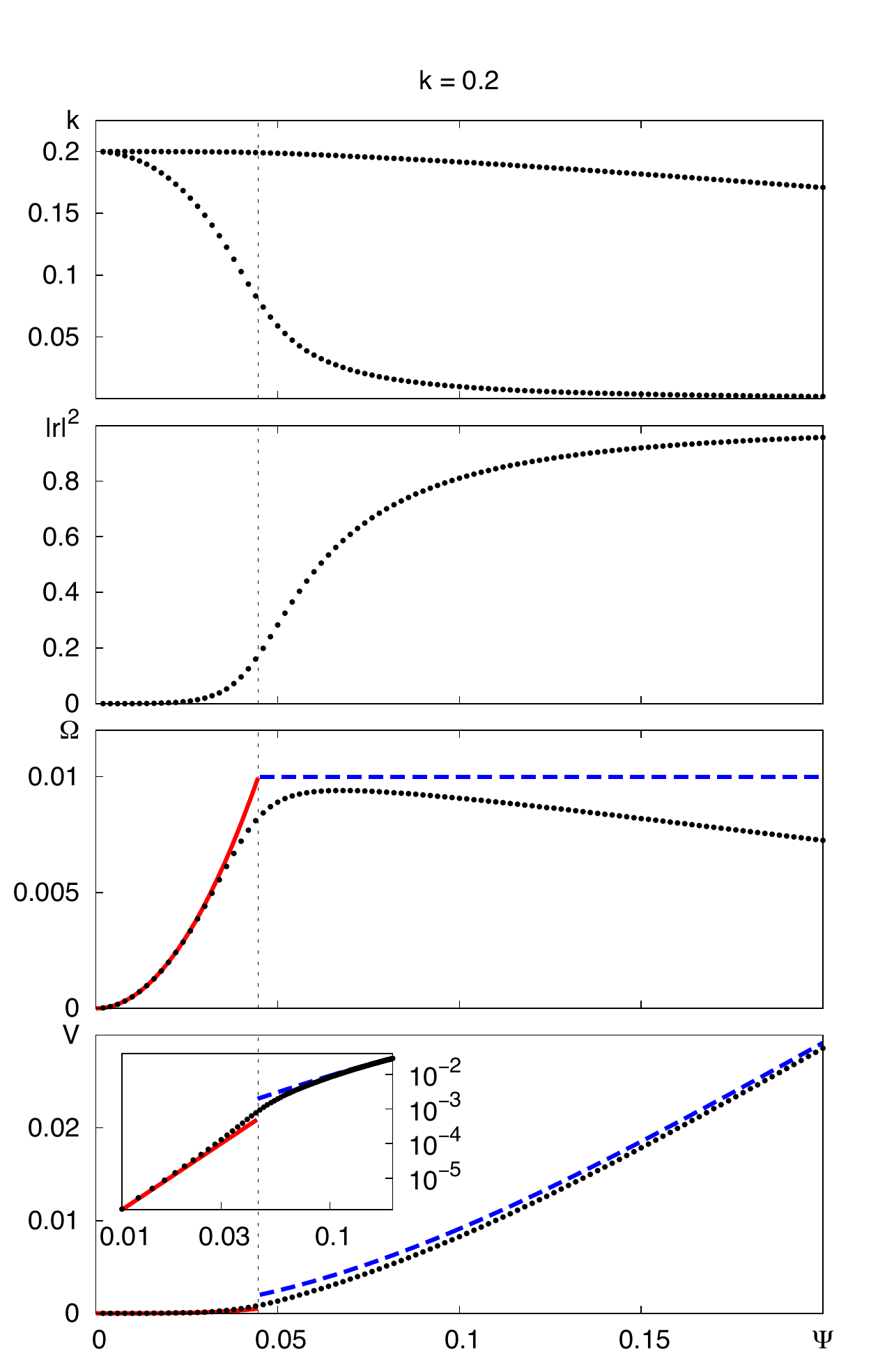}
\caption{Theory for soft magnons, $k = 0.2$ in the lab frame. Top to bottom: wavenumbers in the wall frame $k_\pm$, reflection coefficient $|r|^2$, angular velocity $\Omega$, linear velocity $V$. Inset of the lower panel: $V$ vs $\Psi$ on a log-log scale. Natural units (\ref{eq:units}). Black dots: numerical solution of the equations for steady state (\ref{eq:eom-V-Omega-steady}) with the reactive force and torque (\ref{eq:F-tau}). Red solid line: analytical solution (\ref{eq:Omega-V-soft-redshift}) in the redshift-dominated regime. Blue dashed line:  analytical solution (\ref{eq:Omega-V-soft-reflection}) in the reflection-dominated regime. Vertical dashed lines indicate the crossover (\ref{eq:crossover-soft}) from the redshift-dominated regime on the left to the reflection-dominated regime on the right. }
\label{fig:k-0.2}
\end{figure}

Numerical solutions of the equations of motion (\ref{eq:eom-V-Omega-steady}) for soft spin waves with wavenumber $k = 0.2$ are shown in Fig.~\ref{fig:k-0.2}. For the relaxation time, we used $T = 25$. At low amplitudes, the domain wall precesses slowly, the reflection is weak and the primary mechanism of propulsion is redshift. Neglecting the Doppler shift between the lab and the slowly moving domain wall, we set $k_- = k$. The redshift $\Delta k = k_- - k_+ \ll k$. The reaction force (\ref{eq:F-tau}a) is 
\[
F \approx |\Psi|^2 k \Delta k 
	\approx |\Psi|^2 \omega \Delta \omega
	= 2 |\Psi|^2 \omega \Omega
	\approx 2 |\Psi|^2 \Omega,
\]
where we have used Eq.~(\ref{eq:redshift}) and set $\omega \approx 1$ for soft magnons. The reaction torque (\ref{eq:F-tau}b) is $\tau \approx 2 |\Psi|^2 k$. Substituting these into the steady-state equations (\ref{eq:eom-V-Omega-steady}) yields, in natural units (\ref{eq:units}),
\begin{equation}
\Omega \approx k T |\Psi|^2,
\quad
V \approx T \Omega |\Psi|^2 \approx K T^2|\Psi|^4.
\label{eq:Omega-V-soft-redshift}
\end{equation}

As the amplitude of the incident spin wave $\Psi$ grows, the wall precesses faster and the redshift $\Delta \omega = 2 \Omega$ increases. When $\Delta k$ becomes comparable to $k$, the contribution of reflection becomes comparable to that of redshift---see Eq.~(\ref{eq:r-soft}). For large enough $\Psi$ reflection becomes the dominant force. Assuming perfect reflection, $|r|^2 = 1$, we obtain from Eq.~(\ref{eq:F}) that $F \approx 2 |\Psi|^2 k_-^2$, where $k_-$ is the wavenumber measured in the wall frame. If the incident spin wave has wavenumber $k$ in the lab frame, the Lorentz transformation (\ref{eq:Lorentz-transform}) gives $k_- \approx k - V$ to the first order in $V$ for a soft magnon ($k \ll 1$, $\omega \approx 1$). In essence, this is a Galilean transformation for linear momentum of a massive non-relativistic particle. Eq.~(\ref{eq:eom-V-Omega-steady}) then yields, in natural units (\ref{eq:units}), 
\begin{equation}
V = T (k-V)^2 |\Psi|^2.
\label{eq:V-soft-reflection-implicit} 
\end{equation}
This result can also be derived by representing the spin wave as a stream of ``non-relativistic'' magnons with momenta $p = \hbar k$ and mass $m = \hbar/(s \lambda_0)$ emitted at the rate $\nu = j^1/\hbar = A k |\Psi|^2/\hbar$ and bouncing elastically off the domain wall. (We assume that magnons have a small mass relative to that of the domain wall. This assumption is justified for an antiferromagnet with large classical spins, $S \gg 1$---see Appendix \ref{app:mass-ratio}.)

The velocity of the domain wall grows as $V \approx T k^2 |\Psi|^2$ for small amplitudes $\Psi$ and saturates at $V \sim k$, the group velocity of magnons, when $|\Psi|^2 \gg 1/(kT)$. Generally in reflection dominated regime $|\Psi|^2 \gtrsim 1/(k T)$,
\begin{subequations}
\begin{equation}
V = k \, u(2kT |\Psi|^2),
\quad 
u(\xi) = 1 + \xi^{-1} - \sqrt{2\xi^{-1} + \xi^{-2}}.
\label{eq:xi}
\end{equation}
This result is similar to Eq.~(12) of \textcite{PhysRevLett.112.147204} in the ``non-relativistic'' limit, $V \ll 1$. See Appendix \ref{app:particles} for detailed comparisons. An upper-bound for the angular velocity of the wall can be found directly from Eq.~(\ref{eq:redshift}):
\begin{equation}
\Omega = (\omega_- - \omega_+)/2 < (\omega_- - 1)/2 < k_-^2/4 < k^2/4. 
\label{eq:Omega-soft-reflection}
\end{equation}
\label{eq:Omega-V-soft-reflection}
\end{subequations}
The crossover from the redshift-dominated regim (\ref{eq:Omega-V-soft-redshift}) to the reflection-dominated one (\ref{eq:Omega-V-soft-reflection}) occurs when
\begin{equation}
|\Psi|^2 \sim k/(4T).
\label{eq:crossover-soft}
\end{equation}

\subsection{Hard magnons}

\begin{figure}
\includegraphics[width=0.95\columnwidth]{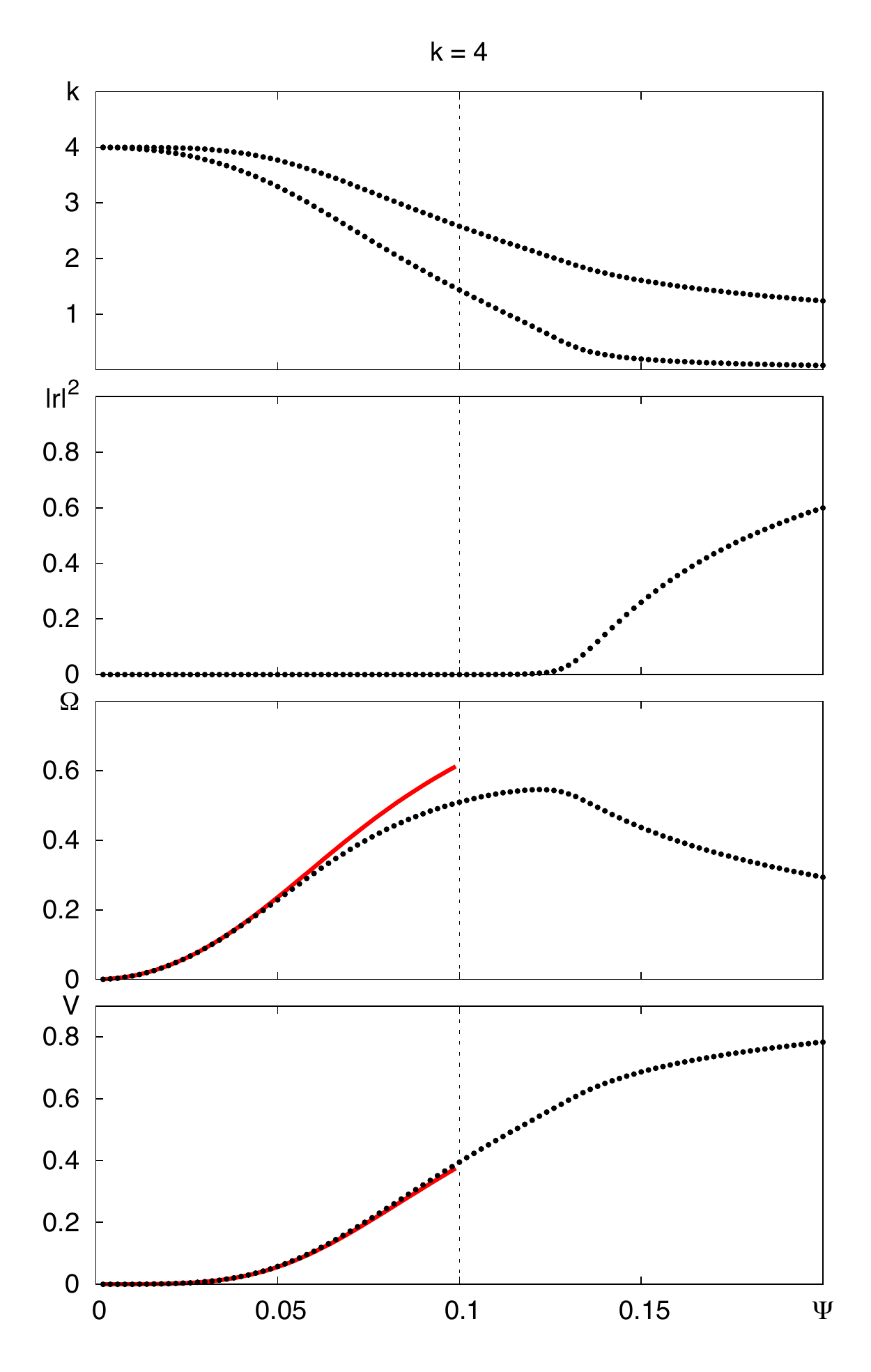}
\caption{Theory for hard magnons, $k = 4$ in the lab frame. Top to bottom: wavenumbers in the wall frame $k_\pm$, reflection coefficient $|r|^2$, angular velocity $\Omega$, linear velocity $V$. Natural units (\ref{eq:units}) are used. Black dots: numerical solution of the equations for steady state (\ref{eq:eom-V-Omega-steady}) with the reactive force and torque (\ref{eq:F-tau}). Red solid line: analytical solution (\ref{eq:Omega-V-hard-redshift}) in the redshift-dominated regime. Vertical dashed lines: the crossover (\ref{eq:crossover-hard}).}
\label{fig:k-4}
\end{figure}

Numerical solutions of the equations of motion (\ref{eq:eom-V-Omega-steady}) for hard spin waves with wavenumber $k = 4$ are shown in Fig.~\ref{fig:k-4}. Again, at small wave amplitudes $\Psi$ redshift dominates (both incident and transmitted magnons are hard in the frame  moving with the wall). As the wave amplitude increases, both the linear and angular velocities initially grow. The growing Doppler shift between the lab and wall frames softens the spin wave in the frame of the wall until the wavevector of the incident wave $k_-$ approaches 1 and the transmitted wave, further redshifted by the precessing wall, becomes soft, $k_+ \ll 1$. At that point, reflection starts to dominate. Weakened transmission translates into a reduced torque and a lower precession frequency. 

In the weak-amplitude regime, where redshift dominates,  
\begin{eqnarray*}
F &\approx& |\Psi|^2 k_- \Delta k \approx |\Psi|^2 \omega_- \Delta \omega = 2 |\Psi|^2 \omega_- \Omega,
\\
\tau &\approx& 2 |\psi|^2 k_-.
\end{eqnarray*}
Transforming from the wall frame to the lab frame,
\[
\omega_- \approx \omega - k V,
\quad
k_- \approx k - \omega V,
\]
and setting $\omega \approx k$ for hard magnons yields 
\begin{subequations}
\begin{eqnarray}
\Omega k (1 - V) |\Psi|^2 &=& V/T,
\\
k(1 - V) |\Psi|^2 &=& \Omega/T,
\end{eqnarray}
\end{subequations}
which can be solved to obtain 
\begin{equation}
V = u(2 k^2 T^2 |\Psi|^4), 
\quad
\Omega = k T |\Psi|^2 (1-V),
\label{eq:Omega-V-hard-redshift}
\end{equation}
where $u(\xi)$ is given in Eq.~(\ref{eq:xi}).

For larger amplitudes, reflection becomes the dominant mechanism. The crossover occurs around
\begin{equation}
|\Psi|^2 \sim 1/(kT).
\label{eq:crossover-hard}
\end{equation}
Waves of large amplitude require a fully ``relativistic'' treatment of the steady-state equation (\ref{eq:eom-V-Omega-steady}).

\subsection{Comparison to numerical simulations}

To check the reliability of our theory, we have conducted numerical simulations of a domain wall in a one-dimensional antiferromagnet with the Hamiltonian
\begin{equation}
H = J \sum_{n=1}^{N-1} \mathbf S_n \cdot \mathbf S_{n+1} - D \sum_{n=1}^N (S^z_n)^2.
\end{equation}
We employed micromagnetic solver OOMMF, which simulates classical dynamics of unit vectors $\mathbf m_n = \mathbf S_n/S$ \cite{oommf}. We used a chain of $N = 4 \times 10^4$ spins with a lattice constant $a = 0.5$ nm. The parameters of the microscopic model and of the field theory are related as follows:
\begin{equation}
\mathcal J = \frac{\hbar S}{2a},
\ 
\rho = \frac{\hbar^2}{4 a J},
\ 
\mathcal M = \frac{\mu}{a},
\ 
A = JS^2 a, 
\ 
K_0 = \frac{2DS^2}{a}.
\end{equation}
We set $\mathcal M = 2.5 \times 10^{-14} $ A$\,$m, $A = 1.25 \times 10^{-31}$ J$\,$m, and $K = 1.25 \times 10^{-17}$ J/m. The gyromagnetic ratio was set at $\gamma = 2.211 \times 10^5$ m/A$\,$s. These coupling constants give natural units (\ref{eq:units})  
\begin{equation}
\lambda_0 = 100 \mbox{ nm},
\quad
t_0 = 28.4 \mbox{ ps},
\quad
s = 3,520 \mbox{ m/s}.
\end{equation}
The Gilbert damping constant $\alpha = 10^{-4}$ gives a relaxation time (\ref{eq:relaxation-time}) $T = 25 t_0 = 710$ ps. Spin waves were excited by driving the leftmost spin with a strong external magnetic field precessing at a fixed angle around the $z$ axis. To prevent reflection of the spin wave from the right end, the Gilbert damping parameter was made spatially inhomogeneous, increasing gradually near the right end of the chain. 

\begin{figure}
\includegraphics[width=0.95\columnwidth]{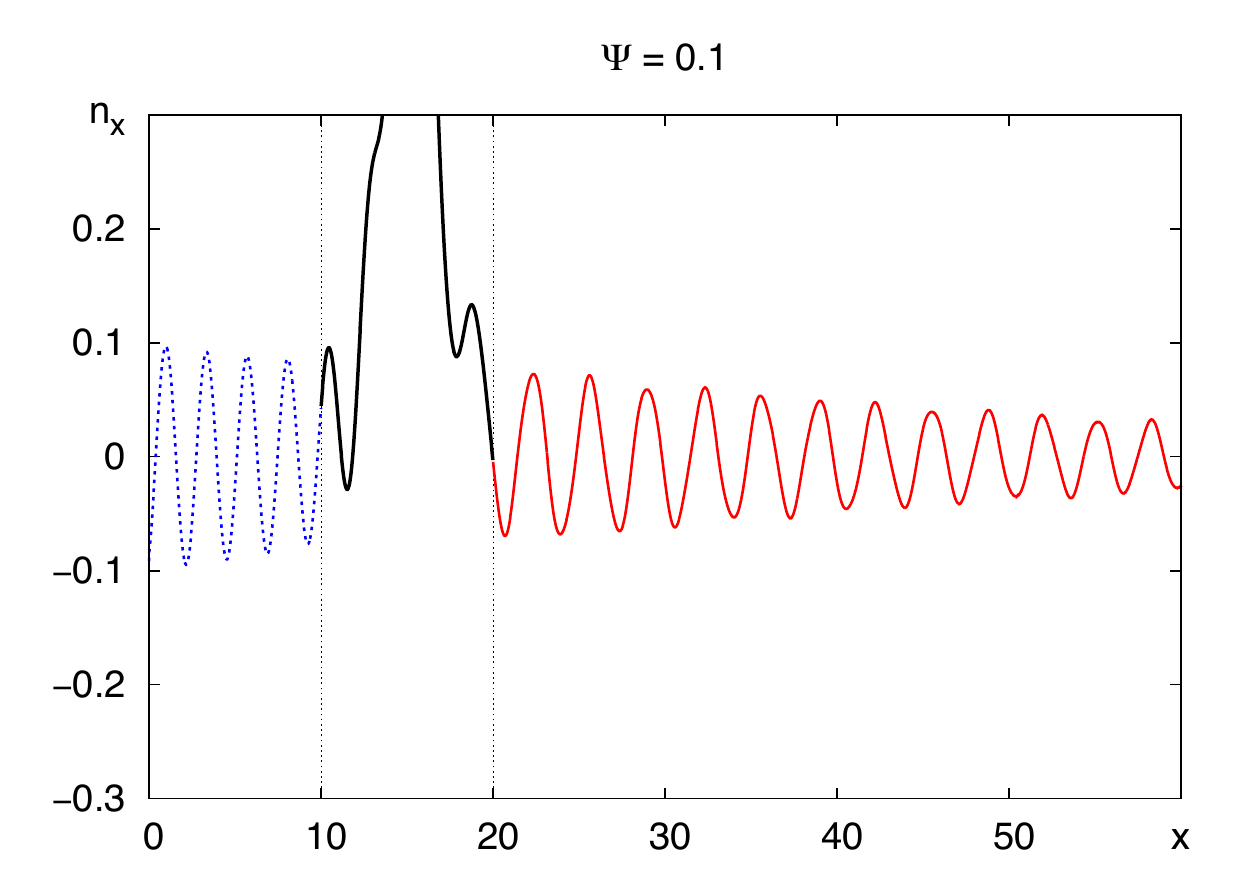}
\includegraphics[width=0.95\columnwidth]{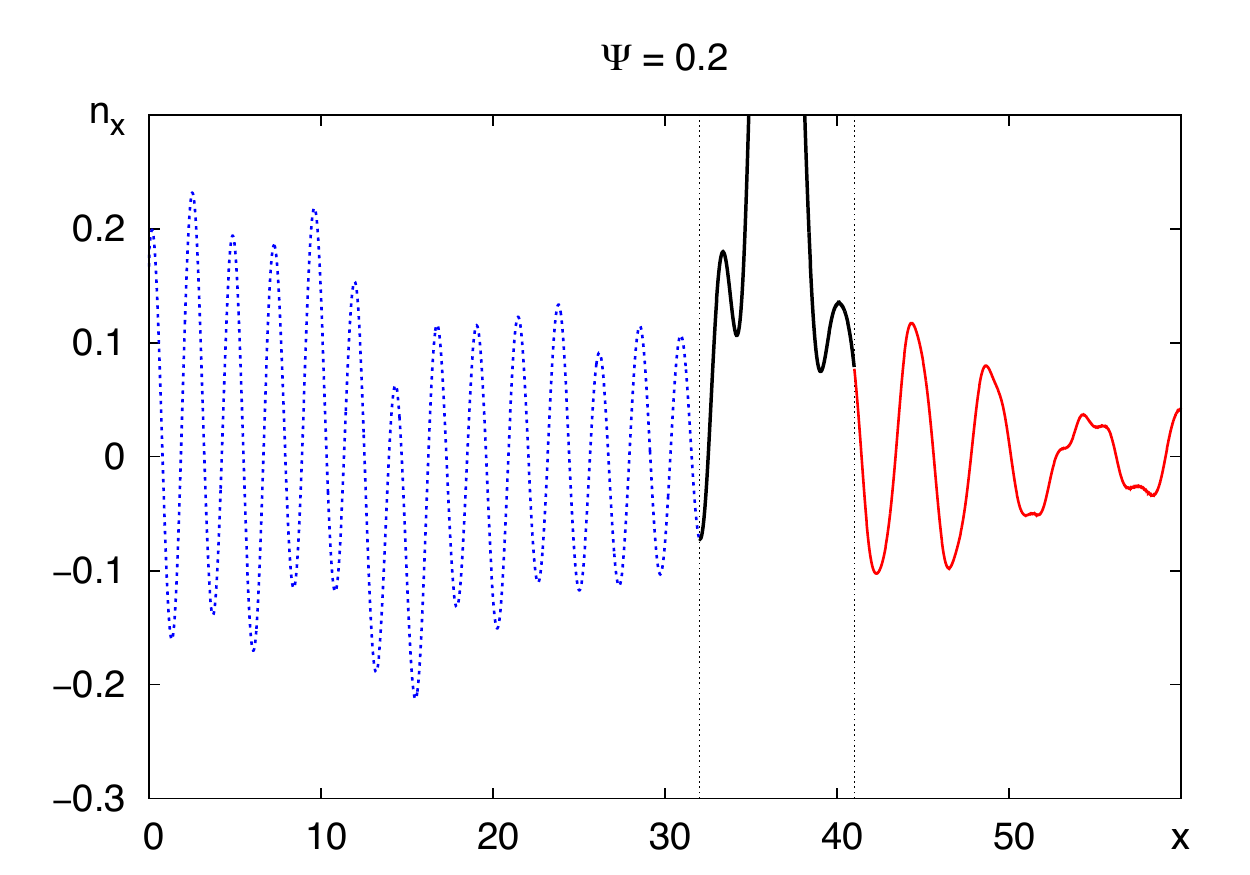}
\caption{Snapshots of the $n_x$ component of the staggered magnetization in simulations with wavenumber $k = 2.66$ and wave amplitudes at the source $\Psi = 0.1$ (top) and 0.2 (bottom). Vertical dotted lines delineate the domain wall.}
\label{fig:snapshots}
\end{figure}

Figure \ref{fig:snapshots} shows snapshots of the $n_x$ component of the staggered magnetization for hard spin waves with $k = 2.66$ incident from the left. Waves with amplitude $\Psi = 0.1$ are sinusoidal, with small damping, on both sides of the domain wall. No reflection is visible on the left-hand side of the domain wall (note the lack of interference). Transmitted waves are clearly redshifted. On the other hand, waves of a larger amplitude $\Psi = 0.2$ generate a strong distortion of the transmitted wave, indicating some sort of structural instability of the domain wall. 

Precession frequency $\Omega$ and linear velocity $V$ of the domain wall observed in the simulations as a function of the spin-wave amplitude $\Psi$ are shown in Fig.~\ref{fig:comparison} along with our theoretical results. We generally find excellent agreement between the two at low spin-wave amplitudes and some deviations when the amplitude increases. 

For soft spin waves ($k = 0.538$), the deviation is due to multiple reflections from the domain wall and the left end. In the reflection-dominated regime ($|\Psi| \gtrsim \sqrt{k/(4T)} = 0.073$), the reflected spin wave bounces off the left end of the chain and exerts additional force and torque on the domain wall. For medium and hard spin waves ($k = 1.718$ and 4.14), deviations occur already in the redshift-dominated regime, where reflection is insignificant. These deviations appear to be caused by the structural instability of the domain wall that sets in at large force and torque. The nature of this instability is not clear and deserves a separate investigation.

\section{Discussion}
\label{sec:discussion}

We have presented a theory of domain-wall propulsion by circularly polarized spin waves in an easy-axis antiferromagnet. A spin wave incident upon a static domain wall (precession frequency $\Omega = 0$, velocity $V = 0$) exerts no force on the wall thanks to the perfect transmission of magnons. However, the inversion of spin carried by transmitted magnons creates a reaction torque, causing the domain wall to precess ($\Omega \neq 0$). The precession creates two effects responsible for the propulsion of the domain wall. 

The first mechanism, identified previously by \textcite{PhysRevLett.112.147204}, is the reflection of spin waves by a precessing domain wall; the reversal of magnon momenta generates a reaction force on the domain wall. We have been able to quantify this effect for the first time by finding an exact solution for a small-amplitude spin wave in the background of a rotating domain wall. We have also identified a second, hitherto unknown mechanism of propulsion, which we termed redshift: the frequency of magnons transmitted through a precessing domain wall is reduced by $2\Omega$; their momenta are reduced as well, resulting in a reaction force on the wall. 

We have obtained closed-form expressions for the net force and torque on the domain wall in its rest frame. The wavenumber and frequency of the incident wave in the wall frame are related to their values in the lab frame by the Lorentz transformation, a symmetry of the antiferromagnet in the continuum limit on length scale larger than the lattice constant. The force and torque balance equations, obtained from the conservation of linear and angular momentum, incorporate the effects of dissipation due to Gilbert damping. The resulting algebraic equations for the velocity of the wall $V$ and its precession frequency $\Omega$ can be readily solved numerically and admit simple approximate solutions in the limits where the propulsion is dominated either by redshift or by reflection. 

\textcite{PhysRevLett.112.147204} showed that translational motion of a domain wall can be also induced by linearly polarized spin waves. But a driving-mechanism is different from circularly polarized cases. Because right- and left-circularly polarized magnons are equally populated in linearly polarized spin waves, a reactive torque on a domain wall vanishes. As a result, the domain wall does not precess $\Omega = 0$ and a reactive force also vanishes. Instead of a reactive force rooted in conservation of linear momentum, a viscous force due to damping of spin waves drives a domain wall toward the source of spin waves. For small Gilbert damping, $\alpha \ll 1$, circularly polarized spin waves induce order-of-magnitude faster motion of a domain wall than linearly polarized ones. 

\section*{Acknowledgments}

We thank Ari Turner for useful discussions and acknowledge the hospitality of Kavli Institute for Theoretical Physics and Aspen Center for Physics, where part of this work has been carried out. OT thanks his students in Quantum Mechanics II for inspiring some of the technical ideas. This research was supported in part by the National Science Foundation under Grants No. DMR-1104753 (JHU), DMR-0840965 (UCLA), PHY-1125915 (KITP), and PHY-1066293 (ACP).

\appendix

\section{Lagrangian dynamics of magnetization}
\label{app:dynamics-general}

\subsection{Ferromagnet}

We begin by reviewing Lagrangian dynamics of magnetization. In a ferromagnet well below the ordering temperature, the magnetization length $\mathcal M$ is fixed and its orientation is represented by the unit-vector field $\mathbf m(\mathbf r)$. Its dynamics is governed by the Landau-Lifshitz equation,
\begin{equation}
\dot{\mathbf m} = \gamma \mathbf h \times \mathbf m + \alpha \mathbf m \times \dot{\mathbf m},
\label{eq:LLG}
\end{equation}
where $\gamma$ is the gyromagnetic ratio, $\mathbf h = - \mathcal M^{-1} \delta U/\delta \mathbf m$ is an effective magnetic field obtained from the potential energy functional $U[\mathbf m(\mathbf r)]$, and $\alpha \ll 1$ is Gilbert's damping constant \cite{Gilbert}. Eq.~(\ref{eq:LLG}) can be obtained from the Lagrangian \cite{Altland-Simons}
\begin{equation}
L = \mathcal J \int \mathbf a(\mathbf m) \cdot \dot{\mathbf m} \, dV - U,
\quad 
\nabla_\mathbf m \times \mathbf a = \mathbf m.
\end{equation}
The first term represents the Berry phase of precessing spins and contains the vector potential $\mathbf a(\mathbf m)$ of a magnetic monopole. $\mathcal J = \mathcal M/\gamma$ is the density of angular momentum. In the Lagrangian formalism, viscous losses are represented by the Rayleigh dissipation function \cite{LL1}
\begin{equation}
R = \frac{\alpha \mathcal J}{2} \int |\dot{\mathbf m}|^2 \, dV.
\label{eq:R-ferro}
\end{equation}

\subsection{N{\'e}el antiferromagnet}
\label{app:Neel}

In a simple antiferromagnet in the absence of an external magnetic field, sublattice magnetizations $\mathbf m_1(\mathbf r)$ and $\mathbf m_2(\mathbf r)$ are nearly antiparallel. Staggered magnetization $\mathbf n = (\mathbf m_1 - \mathbf m_2)/2$ greatly exceeds uniform one $\mathbf m = \mathbf m_1 + \mathbf m_2$. Constraints $|\mathbf m_1|^2 = |\mathbf m_2|^2 = 1$ become 
\begin{equation}
|\mathbf n|^2 = 1, 
\quad
\mathbf m \cdot \mathbf n = 0.
\label{eq:constraints}
\end{equation}
The Lagrangian is
\begin{equation}
L = \mathcal J \int 
		[\mathbf a_1(\mathbf m_1) \cdot \dot{\mathbf m_1} + \mathbf a_2(\mathbf m_2) \cdot \dot{\mathbf m_2}] \, dV 
	- U[\mathbf m_1, \mathbf m_2],
\label{eq:L-m1-m2}
\end{equation}
where $\mathcal J$ is the density of angular momentum on one sublattice. It is convenient to choose different gauges for the vector potentials of the two sublattices: $\mathbf a_1(\mathbf m) = \mathbf a(\mathbf m)$ and $\mathbf a_2(\mathbf m) = \mathbf a(- \mathbf m)$. In the absence of uniform magnetization, the sublattice magnetizations are exactly antiparallel, $\mathbf m_1 = - \mathbf m_2$, and their Berry phases cancel each other out in this gauge. The lowest non-vanishing kinetic contribution to $L$ arises from expanding the Berry-phase terms in Eq.~(\ref{eq:L-m1-m2}) to the first order in $\mathbf m$. Potential energy is also expanded to the lowest order in $\mathbf m$, $U[\mathbf m, \mathbf n] = U[\mathbf n] + \int \left(|\mathbf m|^2/2\chi \right)\, dV$, where $\chi>0$ is proportional to magnetic susceptibility. This results in an effective Lagrangian 
\begin{equation}
L[\mathbf m, \mathbf n] = \int 
	\left[
		\mathcal J \dot{\mathbf n} \cdot (\mathbf n \times \mathbf m) - |\mathbf m|^2/2\chi 
	\right]  \, dV
	- U[\mathbf n]. 
\label{eq:L-m-n}
\end{equation}
With no $\dot{\mathbf m}$ terms in the Lagrangian, uniform magnetization is not a dynamical field but is rather a slave that follows the dynamics of staggered magnetization: 
\begin{equation}
\mathbf m = \mathcal J \chi \, \dot{\mathbf n} \times \mathbf n. 
\label{eq:m-slave}
\end{equation}
Upon eliminating $\mathbf m$, we obtain an effective Lagrangian for staggered magnetization,
\begin{equation}
L[\mathbf n] =  \frac{\rho}{2} \int |\dot{\mathbf n}|^2 \, dV - U[\mathbf n], 
\label{eq:L-n}
\end{equation}
where $\rho = \mathcal J^2 \chi$ quantifies inertia of staggered magnetization. The Rayleigh dissipation function of an antiferromagnet is 
\begin{equation}
R =\alpha \mathcal J \int |\dot{\mathbf n}|^2 \, dV.
\label{eq:R-antiferro}
\end{equation}
We neglect the contribution of uniform magnetization $\mathbf m$ to dissipation, which is proportional to $|\ddot{\mathbf n}|^2$ from Eq.~(\ref{eq:m-slave}), by focusing on slow dynamics. An extra factor of 2, compared to Eq.~(\ref{eq:R-ferro}), reflects the number of sublattices.

In the collective-coordinate approach \cite{Tretiakov2008, Tretiakov2013}, details of a magnetic texture are encoded by a set of generalized coordinates $\mathbf q \equiv \{q_1, q_2, \ldots\}$. The change of $\mathbf n$ with time comes through the time evolution of collective coordinates: $\dot{\mathbf n} = \dot{q}_i \partial \mathbf n/\partial q_i$. We may therefore express the kinetic energy of staggered magnetization as the kinetic energy of collective coordinates: 
\begin{equation}
\frac{1}{2} \int |\dot{\mathbf n}|^2 \, dV = \frac{1}{2} M_{ij} \dot{q}_i \dot{q}_j,
\end{equation}
where the mass tensor is
\begin{equation}
M_{ij} = \rho \int \frac{\partial \mathbf n}{\partial q_i} \cdot \frac{\partial \mathbf n}{\partial q_j} \, dV.
\label{eq:M}
\end{equation}

In a similar way, the Rayleigh dissipation function of staggered magnetization (\ref{eq:R-antiferro}) is represented as
\begin{equation}
\alpha \mathcal J \int |\dot{\mathbf n}|^2 \, dV = \frac{1}{2} D_{ij} \dot{q}_i \dot{q}_j,
\end{equation}
where 
\begin{equation}
D_{ij} = 2 \alpha \mathcal J \int \frac{\partial \mathbf n}{\partial q_i} \cdot \frac{\partial \mathbf n}{\partial q_j} \, dV.
\label{eq:D}
\end{equation}
Clearly, the two tensors are proportional to each other: 
\begin{equation}
D_{ij} = M_{ij}/T.
\label{eq:D-M-relation}
\end{equation}
The proportionality constant
\begin{equation}
T = \frac{\rho}{2 \alpha \mathcal J}
\label{eq:relaxation-time}
\end{equation}
is a relaxation time.

\section{Reflection and transmission amplitudes}
\label{app:r-t}

We discuss the reflection and transmission amplitudes for the SUSY partner Hamiltonians 
\begin{eqnarray}
\mathcal H_0 &=& - d^2/dx^2 + 1 + 2 \omega \Omega \tanh{x},
\\
\mathcal H_1 &=& - d^2/dx^2 + 1 - 2 \mathop{\mathrm{sech}^2}{x} + 2 \omega \Omega \tanh{x},
\end{eqnarray}
where 
\begin{equation}
a= d/dx + \tanh{x} + \omega \Omega, 
\ 
a^\dagger = -d/dx + \tanh{x} + \omega \Omega.
\end{equation}

Far away from the origin, 
\begin{equation}
\psi_0(x) \sim 
	\left\{
		\begin{array}{ll}
			e^{ik_-x} + r_0 e^{-ik_-x} & x \to -\infty,\\
			t_0 e^{ik_+x} & x \to +\infty,
		\end{array}
	\right.
\end{equation}
which defines the reflection and transmission amplitudes $r_0$ and $t_0$ of $\mathcal H_0$. The wavenumbers $k_-$ and $k_+$ are not the same because of the potential step: 
\begin{equation}
k_{\pm}^2 = \omega^2 \mp 2 \omega \Omega - 1.
\label{eq:k-omega-Omega}
\end{equation}

The partner eigenfunction $\psi_1$ can be obtained by the application of the raising operator, $\psi_1 = a^\dagger \psi_0$. It has the following asymptotic form for $x \to -\infty$:  
\[
\psi_1(x) \sim (-1 + \omega \Omega - ik_-) e^{ik_-x} + (-1 + \omega \Omega + ik_-) r_0 e^{-ik_-x}. 
\]
Hence the reflection amplitude 
\begin{equation}
r_1 = r_0 \, \frac{-1 + \omega \Omega - ik_-}{-1 + \omega \Omega + ik_-}.
\end{equation}
Clearly, $|r_1| = |r_0|$, so we will refer to both of these as simply $|r|$.

Along the same lines, we obtain
\begin{equation}
t_1 = t_0 \, \frac{+1 + \omega \Omega - ik_+}{-1 + \omega \Omega + ik_-}.
\end{equation}
It can be checked, with the aid of Eq.~(\ref{eq:k-omega-Omega}), that again $|t_1| = |t_0| = |t|$.

The reflection amplitude for the $\tanh{x}$ potential is known \cite{LL3, JHEP.2011.73}:
\begin{equation}
|r|^2 = \frac{\sinh^2{[\frac{\pi}{2}(k_+ - k_-)]}}{\sinh^2{[\frac{\pi}{2}(k_+ + k_-)]}}.
\end{equation}
The transmission amplitude is related to the reflection amplitude in the usual way \cite{LL3},
\begin{equation}
|r|^2 k_- + |t|^2 k_+ = k_-.
\end{equation}

\section{Mass of a magnon vs mass of a domain wall}
\label{app:mass-ratio}

Here we compute the ratio of the magnon mass $m$ to the domain-wall mass $M$, or equivalently, the ratio of their rest energies.

The energy density (\ref{eq:def-T}) of a circularly polarized spin wave $\psi(x, t) = \Psi e^{-i \omega t + i k x}$ (\ref{eq:plane-wave}) in the background of a N{\'e}el ground state (\ref{eq:Neel-states}) is
\begin{equation}
s^2 T^{00} = \rho \omega^2 |\Psi|^2.
\end{equation}
The number density of magnons is given by the absolute value of the spin density $j^0$~(\ref{eq:j-uniform}) divided by the spin of each magnon $\hbar$. The energy of one magnon is
\begin{equation}
E = \frac{s^2 T^{00}}{|j^0 / \hbar|} = \frac{\rho \omega^2 |\Psi|^2}{\rho |\omega| |\Psi|^2 / \hbar} = \hbar |\omega|,
\end{equation}
which satisfies a ``relativistic'' energy-momentum relation
\begin{equation}
E^2 = p^2 s^2 + (\hbar \Delta)^2,
\end{equation}
where $p = \hbar k$ is the momentum of a magnon and $\Delta = \sqrt{K_0 / \rho} = 1/t_0$ is the frequency gap. The rest energy of a magnon is $m s^2 = \hbar \Delta$. The rest energy of the wall (\ref{eq:domain-wall-static}) is
$M s^2 = 2 \sqrt{A K_0}$. Their ratio is
\[
\frac{M}{m} = \frac{M s^2}{m s^2} = \frac{2 \sqrt{A K_0}}{\hbar \Delta} = \frac{2 \mathcal{J} \sqrt{A \chi}}{\hbar}.
\]

A one-dimensional antiferromagnet with the Hamiltonian
\begin{equation}
H = J \sum_{n=1}^{N-1} \mathbf S_n \cdot \mathbf S_{n+1} - D \sum_{n=1}^N (S^z_n)^2
\end{equation}
has the following continuum parameters: scalar density of spin angular momentum $\mathcal J = \hbar S /2 a$, exchange constant $A = J S^2 a$, and susceptibility $\chi = a / J S^2$, where $a$ is the lattice constant and $S$ is the length of a spin at each site. Then the ratio becomes
\begin{equation}
M/m = S.
\end{equation}
The domain wall is much heavier than a magnon, $M \gg m$, in the classical limit $S \gg 1$. 

We note that a rescaled mass-ratio $2 m / \hbar M = \left( \mathcal J \sqrt{A \chi} \right)^{-1}$ is the coupling constant $g = 2 / \hbar S$ in Haldane's Lagrangian density \cite{PhysRevLett.50.1153}
\begin{equation}
\mathcal L = \frac{|\dot{\mathbf n}|^2 - s^2 |\mathbf n'|^2 - \Delta^2 (\hat{\mathbf z} \times \mathbf n)^2}{2 gs}.
\end{equation}

\section{Reactive force from magnon reflection}
\label{app:particles}

An intuitive way to derive the spin-wave force on the domain wall is to picture the spin wave as a flux of magnons, particles carrying angular momentum $\hbar$, momentum $\hbar k$, and energy $\hbar \omega$. Their velocity
\begin{equation}
v = d(\hbar \omega)/d(\hbar k) = s^2 k/\omega
\end{equation}
equals the group velocity of spin waves. With the spin current (\ref{eq:j-uniform}), the rate at which magnons are emitted by the source is 
\begin{equation}
\nu = j^1/\hbar = \hbar^{-1} A |\Psi|^2 k. 
\label{eq:nu}
\end{equation}
The rate $\nu'$ at which magnons hit the domain wall is reduced if the wall is moving: 
\begin{equation}
\nu' = \nu(1 - V/v) = \hbar^{-1} A |\Psi|^2(k - \omega V/s^2).
\label{eq:nu'}
\end{equation}

Consider an elastic collision of a magnon with a domain wall. In the reference frame moving with the wall, the magnon's initial momentum and energy are
\[
\hbar k_- = \frac{\hbar(k - \omega V/s^2)}{\sqrt{1-V^2/s^2}},
\quad
\hbar \omega_- = \frac{\hbar(\omega - kV)}{\sqrt{1-V^2/s^2}}.
\]
After the collision, it is reversed to $- \hbar k_-$, assuming that the mass of the domain wall $M$ greatly exceeds the mass $m$ of the magnon. The final momentum of the magnon in the lab frame is
\[
\hbar k' = \frac{-\hbar k_- + \omega_- V/s^2}{\sqrt{1-V^2/s^2}}
	= \frac{2\hbar \omega V/s^2 - \hbar k(1 + V^2/s^2)}{1-V^2 s^2}.
\]
Momentum transferred to the wall in the collision is 
\begin{equation}
\Delta p = \hbar k - \hbar k' = \frac{2\hbar(k - \omega V/s^2)}{1 - V^2/s^2}.
\end{equation}
The time-averaged force is 
\begin{equation}
\bar{F} = \nu' \Delta p 
	= \frac{2 A |\Psi|^2 (k - \omega V/s^2)^2}{1 - V^2/s^2}
	= 2 A |\Psi|^2 k_-^2.
\label{eq:F-magnons}
\end{equation}
This expression agrees with our Eq.~(\ref{eq:F}) for perfect reflection, $|r|^2 = 1$. The result of \textcite{PhysRevLett.112.147204} is $\bar{F} = 2 A |\Psi|^2 k (k - \omega V)$, which is different from ours (\ref{eq:F-magnons}).

Alternatively, we may obtain the force in the frame of the wall and use the convenient fact that the force is the same in both frames in (1+1)-dimensional relativity (in 3+1 dimensions, it is the longitudinal component of the force that remains the same in both frames \cite{Steane}). In the wall frame, magnons are emitted and collide with the wall at the same rate 
\[
\nu = j^1/\hbar = \hbar^{-1} A |\Psi|^2 k_-.
\]
Momentum transfer in an elastic collision with a stationary wall is $\Delta p = 2 \hbar k_-$, which yields the average force
\begin{equation}
\bar{F} = \nu \Delta p = 2 A |\Psi|^2 k_-^2,
\end{equation}
in agreement with our previous result (\ref{eq:F-magnons}).

\bibliographystyle{apsrev4-1}
\bibliography{AF-wall-magnons}

\begin{thebibliography}{23}%
\makeatletter
\providecommand \@ifxundefined [1]{%
 \@ifx{#1\undefined}
}%
\providecommand \@ifnum [1]{%
 \ifnum #1\expandafter \@firstoftwo
 \else \expandafter \@secondoftwo
 \fi
}%
\providecommand \@ifx [1]{%
 \ifx #1\expandafter \@firstoftwo
 \else \expandafter \@secondoftwo
 \fi
}%
\providecommand \natexlab [1]{#1}%
\providecommand \enquote  [1]{``#1''}%
\providecommand \bibnamefont  [1]{#1}%
\providecommand \bibfnamefont [1]{#1}%
\providecommand \citenamefont [1]{#1}%
\providecommand \href@noop [0]{\@secondoftwo}%
\providecommand \href [0]{\begingroup \@sanitize@url \@href}%
\providecommand \@href[1]{\@@startlink{#1}\@@href}%
\providecommand \@@href[1]{\endgroup#1\@@endlink}%
\providecommand \@sanitize@url [0]{\catcode `\\12\catcode `\$12\catcode
  `\&12\catcode `\#12\catcode `\^12\catcode `\_12\catcode `\%12\relax}%
\providecommand \@@startlink[1]{}%
\providecommand \@@endlink[0]{}%
\providecommand \url  [0]{\begingroup\@sanitize@url \@url }%
\providecommand \@url [1]{\endgroup\@href {#1}{\urlprefix }}%
\providecommand \urlprefix  [0]{URL }%
\providecommand \Eprint [0]{\href }%
\providecommand \doibase [0]{http://dx.doi.org/}%
\providecommand \selectlanguage [0]{\@gobble}%
\providecommand \bibinfo  [0]{\@secondoftwo}%
\providecommand \bibfield  [0]{\@secondoftwo}%
\providecommand \translation [1]{[#1]}%
\providecommand \BibitemOpen [0]{}%
\providecommand \bibitemStop [0]{}%
\providecommand \bibitemNoStop [0]{.\EOS\space}%
\providecommand \EOS [0]{\spacefactor3000\relax}%
\providecommand \BibitemShut  [1]{\csname bibitem#1\endcsname}%
\let\auto@bib@innerbib\@empty
\bibitem [{\citenamefont {Parkin}\ \emph {et~al.}(2008)\citenamefont {Parkin},
  \citenamefont {Hayashi},\ and\ \citenamefont {Thomas}}]{Science.320.190}%
  \BibitemOpen
  \bibfield  {author} {\bibinfo {author} {\bibfnamefont {S.~S.~P.}\
  \bibnamefont {Parkin}}, \bibinfo {author} {\bibfnamefont {M.}~\bibnamefont
  {Hayashi}}, \ and\ \bibinfo {author} {\bibfnamefont {L.}~\bibnamefont
  {Thomas}},\ }\href {\doibase 10.1126/science.1145799} {\bibfield  {journal}
  {\bibinfo  {journal} {Science}\ }\textbf {\bibinfo {volume} {320}},\ \bibinfo
  {pages} {190} (\bibinfo {year} {2008})}\BibitemShut {NoStop}%
\bibitem [{\citenamefont {Berger}(1978)}]{Berger1978}%
  \BibitemOpen
  \bibfield  {author} {\bibinfo {author} {\bibfnamefont {L.}~\bibnamefont
  {Berger}},\ }\href {\doibase http://dx.doi.org/10.1063/1.324716} {\bibfield
  {journal} {\bibinfo  {journal} {J. App. Phys.}\ }\textbf {\bibinfo {volume}
  {49}},\ \bibinfo {pages} {2156} (\bibinfo {year} {1978})}\BibitemShut
  {NoStop}%
\bibitem [{\citenamefont {Slonczewski}(1996)}]{Slonczewski1996}%
  \BibitemOpen
  \bibfield  {author} {\bibinfo {author} {\bibfnamefont {J.}~\bibnamefont
  {Slonczewski}},\ }\href {\doibase 10.1016/0304-8853(96)00062-5} {\bibfield
  {journal} {\bibinfo  {journal} {J. Mag. Magn. Mater.}\ }\textbf {\bibinfo
  {volume} {159}},\ \bibinfo {pages} {L1} (\bibinfo {year} {1996})}\BibitemShut
  {NoStop}%
\bibitem [{\citenamefont {Berger}(1996)}]{Berger1996}%
  \BibitemOpen
  \bibfield  {author} {\bibinfo {author} {\bibfnamefont {L.}~\bibnamefont
  {Berger}},\ }\href {\doibase 10.1103/PhysRevB.54.9353} {\bibfield  {journal}
  {\bibinfo  {journal} {Phys. Rev. B}\ }\textbf {\bibinfo {volume} {54}},\
  \bibinfo {pages} {9353} (\bibinfo {year} {1996})}\BibitemShut {NoStop}%
\bibitem [{\citenamefont {Schryer}\ and\ \citenamefont
  {Walker}(1974)}]{Walker1973}%
  \BibitemOpen
  \bibfield  {author} {\bibinfo {author} {\bibfnamefont {N.~L.}\ \bibnamefont
  {Schryer}}\ and\ \bibinfo {author} {\bibfnamefont {L.~R.}\ \bibnamefont
  {Walker}},\ }\href {\doibase http://dx.doi.org/10.1063/1.1663252} {\bibfield
  {journal} {\bibinfo  {journal} {J. Appl. Phys}\ }\textbf {\bibinfo {volume}
  {45}},\ \bibinfo {pages} {5406} (\bibinfo {year} {1974})}\BibitemShut
  {NoStop}%
\bibitem [{\citenamefont {Hinzke}\ and\ \citenamefont
  {Nowak}(2011)}]{Hinzke2011}%
  \BibitemOpen
  \bibfield  {author} {\bibinfo {author} {\bibfnamefont {D.}~\bibnamefont
  {Hinzke}}\ and\ \bibinfo {author} {\bibfnamefont {U.}~\bibnamefont {Nowak}},\
  }\href {\doibase 10.1103/PhysRevLett.107.027205} {\bibfield  {journal}
  {\bibinfo  {journal} {Phys. Rev. Lett.}\ }\textbf {\bibinfo {volume} {107}},\
  \bibinfo {pages} {027205} (\bibinfo {year} {2011})}\BibitemShut {NoStop}%
\bibitem [{\citenamefont {Yan}\ \emph {et~al.}(2011)\citenamefont {Yan},
  \citenamefont {Wang},\ and\ \citenamefont {Wang}}]{Yan2011}%
  \BibitemOpen
  \bibfield  {author} {\bibinfo {author} {\bibfnamefont {P.}~\bibnamefont
  {Yan}}, \bibinfo {author} {\bibfnamefont {X.~S.}\ \bibnamefont {Wang}}, \
  and\ \bibinfo {author} {\bibfnamefont {X.~R.}\ \bibnamefont {Wang}},\ }\href
  {\doibase 10.1103/PhysRevLett.107.177207} {\bibfield  {journal} {\bibinfo
  {journal} {Phys. Rev. Lett.}\ }\textbf {\bibinfo {volume} {107}},\ \bibinfo
  {pages} {177207} (\bibinfo {year} {2011})}\BibitemShut {NoStop}%
\bibitem [{\citenamefont {Kovalev}\ and\ \citenamefont
  {Tserkovnyak}(2012)}]{Kovalev2012}%
  \BibitemOpen
  \bibfield  {author} {\bibinfo {author} {\bibfnamefont {A.~A.}\ \bibnamefont
  {Kovalev}}\ and\ \bibinfo {author} {\bibfnamefont {Y.}~\bibnamefont
  {Tserkovnyak}},\ }\href {\doibase 10.1209/0295-5075/97/67002} {\bibfield
  {journal} {\bibinfo  {journal} {EPL}\ }\textbf {\bibinfo {volume} {97}},\
  \bibinfo {pages} {67002} (\bibinfo {year} {2012})}\BibitemShut {NoStop}%
\bibitem [{\citenamefont {Tveten}\ \emph {et~al.}(2014)\citenamefont {Tveten},
  \citenamefont {Qaiumzadeh},\ and\ \citenamefont
  {Brataas}}]{PhysRevLett.112.147204}%
  \BibitemOpen
  \bibfield  {author} {\bibinfo {author} {\bibfnamefont {E.~G.}\ \bibnamefont
  {Tveten}}, \bibinfo {author} {\bibfnamefont {A.}~\bibnamefont {Qaiumzadeh}},
  \ and\ \bibinfo {author} {\bibfnamefont {A.}~\bibnamefont {Brataas}},\ }\href
  {\doibase 10.1103/PhysRevLett.112.147204} {\bibfield  {journal} {\bibinfo
  {journal} {Phys. Rev. Lett.}\ }\textbf {\bibinfo {volume} {112}},\ \bibinfo
  {pages} {147204} (\bibinfo {year} {2014})}\BibitemShut {NoStop}%
\bibitem [{\citenamefont {Tretiakov}\ \emph {et~al.}(2008)\citenamefont
  {Tretiakov}, \citenamefont {Clarke}, \citenamefont {Chern}, \citenamefont
  {Bazaliy},\ and\ \citenamefont {Tchernyshyov}}]{Tretiakov2008}%
  \BibitemOpen
  \bibfield  {author} {\bibinfo {author} {\bibfnamefont {O.~A.}\ \bibnamefont
  {Tretiakov}}, \bibinfo {author} {\bibfnamefont {D.}~\bibnamefont {Clarke}},
  \bibinfo {author} {\bibfnamefont {G.-W.}\ \bibnamefont {Chern}}, \bibinfo
  {author} {\bibfnamefont {Y.~B.}\ \bibnamefont {Bazaliy}}, \ and\ \bibinfo
  {author} {\bibfnamefont {O.}~\bibnamefont {Tchernyshyov}},\ }\href {\doibase
  10.1103/PhysRevLett.100.127204} {\bibfield  {journal} {\bibinfo  {journal}
  {Phys. Rev. Lett.}\ }\textbf {\bibinfo {volume} {100}},\ \bibinfo {pages}
  {127204} (\bibinfo {year} {2008})}\BibitemShut {NoStop}%
\bibitem [{\citenamefont {Tveten}\ \emph {et~al.}(2013)\citenamefont {Tveten},
  \citenamefont {Qaiumzadeh}, \citenamefont {Tretiakov},\ and\ \citenamefont
  {Brataas}}]{Tretiakov2013}%
  \BibitemOpen
  \bibfield  {author} {\bibinfo {author} {\bibfnamefont {E.~G.}\ \bibnamefont
  {Tveten}}, \bibinfo {author} {\bibfnamefont {A.}~\bibnamefont {Qaiumzadeh}},
  \bibinfo {author} {\bibfnamefont {O.~A.}\ \bibnamefont {Tretiakov}}, \ and\
  \bibinfo {author} {\bibfnamefont {A.}~\bibnamefont {Brataas}},\ }\href
  {\doibase 10.1103/PhysRevLett.110.127208} {\bibfield  {journal} {\bibinfo
  {journal} {Phys. Rev. Lett.}\ }\textbf {\bibinfo {volume} {110}},\ \bibinfo
  {pages} {127208} (\bibinfo {year} {2013})}\BibitemShut {NoStop}%
\bibitem [{\citenamefont {Clarke}\ \emph {et~al.}(2008)\citenamefont {Clarke},
  \citenamefont {Tretiakov}, \citenamefont {Chern}, \citenamefont {Bazaliy},\
  and\ \citenamefont {Tchernyshyov}}]{Clarke2008}%
  \BibitemOpen
  \bibfield  {author} {\bibinfo {author} {\bibfnamefont {D.~J.}\ \bibnamefont
  {Clarke}}, \bibinfo {author} {\bibfnamefont {O.~A.}\ \bibnamefont
  {Tretiakov}}, \bibinfo {author} {\bibfnamefont {G.-W.}\ \bibnamefont
  {Chern}}, \bibinfo {author} {\bibfnamefont {Y.~B.}\ \bibnamefont {Bazaliy}},
  \ and\ \bibinfo {author} {\bibfnamefont {O.}~\bibnamefont {Tchernyshyov}},\
  }\href {\doibase 10.1103/PhysRevB.78.134412} {\bibfield  {journal} {\bibinfo
  {journal} {Phys. Rev. B}\ }\textbf {\bibinfo {volume} {78}},\ \bibinfo
  {pages} {134412} (\bibinfo {year} {2008})}\BibitemShut {NoStop}%
\bibitem [{\citenamefont {Sukumar}(1985)}]{JPhysA.18.2917}%
  \BibitemOpen
  \bibfield  {author} {\bibinfo {author} {\bibfnamefont {C.~V.}\ \bibnamefont
  {Sukumar}},\ }\href {\doibase 10.1088/0305-4470/18/15/020} {\bibfield
  {journal} {\bibinfo  {journal} {J. Phys. A}\ }\textbf {\bibinfo {volume}
  {18}},\ \bibinfo {pages} {2917} (\bibinfo {year} {1985})}\BibitemShut
  {NoStop}%
\bibitem [{\citenamefont {Cooper}\ \emph {et~al.}(1995)\citenamefont {Cooper},
  \citenamefont {Khare},\ and\ \citenamefont {Sukhatme}}]{PhysRep.251.267}%
  \BibitemOpen
  \bibfield  {author} {\bibinfo {author} {\bibfnamefont {F.}~\bibnamefont
  {Cooper}}, \bibinfo {author} {\bibfnamefont {A.}~\bibnamefont {Khare}}, \
  and\ \bibinfo {author} {\bibfnamefont {U.}~\bibnamefont {Sukhatme}},\ }\href
  {\doibase 10.1016/0370-1573(94)00080-M} {\bibfield  {journal} {\bibinfo
  {journal} {Phys. Rep.}\ }\textbf {\bibinfo {volume} {251}},\ \bibinfo {pages}
  {267} (\bibinfo {year} {1995})}\BibitemShut {NoStop}%
\bibitem [{\citenamefont {Haldane}(1983)}]{PhysRevLett.50.1153}%
  \BibitemOpen
  \bibfield  {author} {\bibinfo {author} {\bibfnamefont {F.~D.~M.}\
  \bibnamefont {Haldane}},\ }\href {\doibase 10.1103/PhysRevLett.50.1153}
  {\bibfield  {journal} {\bibinfo  {journal} {Phys. Rev. Lett.}\ }\textbf
  {\bibinfo {volume} {50}},\ \bibinfo {pages} {1153} (\bibinfo {year}
  {1983})}\BibitemShut {NoStop}%
\bibitem [{\citenamefont {Donahue}\ and\ \citenamefont {Porter}(1999)}]{oommf}%
  \BibitemOpen
  \bibfield  {author} {\bibinfo {author} {\bibfnamefont {M.}~\bibnamefont
  {Donahue}}\ and\ \bibinfo {author} {\bibfnamefont {D.}~\bibnamefont
  {Porter}},\ }\href@noop {} {\emph {\bibinfo {title} {OOMMF User's Guide,
  Version 1.0}}},\ \bibinfo {address} {NIST, Gaithersburg, MD} (\bibinfo {year}
  {1999})\BibitemShut {NoStop}%
\bibitem [{\citenamefont {Landau}\ and\ \citenamefont
  {Lifshitz}(1976{\natexlab{a}})}]{LL1}%
  \BibitemOpen
  \bibfield  {author} {\bibinfo {author} {\bibfnamefont {L.~D.}\ \bibnamefont
  {Landau}}\ and\ \bibinfo {author} {\bibfnamefont {E.~M.}\ \bibnamefont
  {Lifshitz}},\ }\href@noop {} {\emph {\bibinfo {title} {Mechanics}}},\
  \bibinfo {edition} {3rd}\ ed.\ (\bibinfo  {publisher}
  {Butterworth-Heinemann},\ \bibinfo {address} {Oxford},\ \bibinfo {year}
  {1976})\BibitemShut {NoStop}%
\bibitem [{\citenamefont {P{\"o}schl}\ and\ \citenamefont
  {Teller}(1933)}]{ZPhys.83.143}%
  \BibitemOpen
  \bibfield  {author} {\bibinfo {author} {\bibfnamefont {G.}~\bibnamefont
  {P{\"o}schl}}\ and\ \bibinfo {author} {\bibfnamefont {E.}~\bibnamefont
  {Teller}},\ }\href {\doibase 10.1007/BF01331132} {\bibfield  {journal}
  {\bibinfo  {journal} {Z. Phys.}\ }\textbf {\bibinfo {volume} {83}},\ \bibinfo
  {pages} {143} (\bibinfo {year} {1933})}\BibitemShut {NoStop}%
\bibitem [{\citenamefont {Gilbert}(2004)}]{Gilbert}%
  \BibitemOpen
  \bibfield  {author} {\bibinfo {author} {\bibfnamefont {T.}~\bibnamefont
  {Gilbert}},\ }\href {\doibase 10.1109/TMAG.2004.836740} {\bibfield  {journal}
  {\bibinfo  {journal} {IEEE Trans. Magn.}\ }\textbf {\bibinfo {volume} {40}},\
  \bibinfo {pages} {3443} (\bibinfo {year} {2004})}\BibitemShut {NoStop}%
\bibitem [{\citenamefont {Altland}\ and\ \citenamefont
  {Simons}(2010)}]{Altland-Simons}%
  \BibitemOpen
  \bibfield  {author} {\bibinfo {author} {\bibfnamefont {A.}~\bibnamefont
  {Altland}}\ and\ \bibinfo {author} {\bibfnamefont {B.}~\bibnamefont
  {Simons}},\ }\href@noop {} {\emph {\bibinfo {title} {Condensed matter field
  theory}}},\ \bibinfo {edition} {2nd}\ ed.\ (\bibinfo  {publisher} {Cambridge
  University Press},\ \bibinfo {address} {Cambridge},\ \bibinfo {year}
  {2010})\BibitemShut {NoStop}%
\bibitem [{\citenamefont {Landau}\ and\ \citenamefont
  {Lifshitz}(1976{\natexlab{b}})}]{LL3}%
  \BibitemOpen
  \bibfield  {author} {\bibinfo {author} {\bibfnamefont {L.~D.}\ \bibnamefont
  {Landau}}\ and\ \bibinfo {author} {\bibfnamefont {E.~M.}\ \bibnamefont
  {Lifshitz}},\ }\href@noop {} {\emph {\bibinfo {title} {Quantum Mechanics}}},\
  \bibinfo {edition} {3rd}\ ed.\ (\bibinfo  {publisher}
  {Butterworth-Heinemann},\ \bibinfo {address} {Oxford},\ \bibinfo {year}
  {1976})\BibitemShut {NoStop}%
\bibitem [{\citenamefont {Boonserm}\ and\ \citenamefont
  {Visser}(2011)}]{JHEP.2011.73}%
  \BibitemOpen
  \bibfield  {author} {\bibinfo {author} {\bibfnamefont {P.}~\bibnamefont
  {Boonserm}}\ and\ \bibinfo {author} {\bibfnamefont {M.}~\bibnamefont
  {Visser}},\ }\href {\doibase 10.1007/JHEP03(2011)073} {\bibfield  {journal}
  {\bibinfo  {journal} {JHEP}\ }\textbf {\bibinfo {volume} {2011}},\ \bibinfo
  {pages} {73} (\bibinfo {year} {2011})}\BibitemShut {NoStop}%
\bibitem [{\citenamefont {Steane}(2012)}]{Steane}%
  \BibitemOpen
  \bibfield  {author} {\bibinfo {author} {\bibfnamefont {A.~M.}\ \bibnamefont
  {Steane}},\ }\href@noop {} {\emph {\bibinfo {title} {Relativity made
  relatively easy}}},\ \bibinfo {edition} {1st}\ ed.\ (\bibinfo  {publisher}
  {Oxford University Press},\ \bibinfo {address} {Oxford},\ \bibinfo {year}
  {2012})\BibitemShut {NoStop}%
\end{thebibliography}%

\end{document}